\documentclass[twocolumn,aps,prd,floatfix,preprintnumbers,a4paper,nofootinbib,superscriptaddress,10pt]{revtex4-1} 
\usepackage{lipsum}
\usepackage{multirow}
\usepackage{import}
\usepackage{tabularx}
\usepackage[titletoc,title]{appendix}

\usepackage{graphicx}

\usepackage{amsmath}
\usepackage{amssymb}
\usepackage{amsfonts}
\usepackage{mathtools}
\usepackage{bm}
\usepackage{import}

\usepackage[utf8]{inputenc}
\usepackage{txfonts}

\usepackage[table]{xcolor}
\usepackage{url}
\usepackage[colorlinks, pdfborder={0 0 0}]{hyperref}
\definecolor{LinkColor}{rgb}{0.75, 0, 0}
\definecolor{CiteColor}{rgb}{0.75, 0, 0}
\definecolor{UrlColor}{rgb}{0, 0, 0.75}
\hypersetup{linkcolor=LinkColor}
\hypersetup{citecolor=CiteColor}
\hypersetup{urlcolor=UrlColor}

\usepackage{rotating}
\usepackage{booktabs}

\usepackage{dcolumn}
\newcolumntype{d}{D{.}{.}{-1}}
\newcolumntype{b}[1]{D{(}{(}{#1}}

\newcommand{\Cardiff}{School of Physics and Astronomy, Cardiff University, Queens Buildings, Cardiff, CF24 3AA, United Kingdom}

\newcommand{\Zurich}{Physik-Institut, Universit\"at Z\"urich, Winterthurerstrasse 190, 8057 Z\"urich, Switzerland}
\newcommand{\UIB}{Departament de F\'isica, Universitat de les Illes Balears, IAC3 – IEEC, Crta. Valldemossa km 7.5, E-07122 Palma, Spain}
\newcommand{\Amsterdam}{Institute for High-Energy Physics, University of Amsterdam, Science Park 904, 1098 XH Amsterdam, The Netherlands}
\newcommand{\KCL}{King's  College  London,  Strand,  London  WC2R  2LS,  United Kingdom}
\newcommand{\AEI}{Max Planck Institute for Gravitational Physics (Albert Einstein Institute),
Callinstrasse 38, D-30167 Hannover, Germany}
\newcommand{\Leibniz}{Leibniz Universitaat Hannover, 30167 Hannover, Germany}
\newcommand{\Caltech}{Theoretical Astrophysics Group, California Institute of Technology, Pasadena, CA 91125, U.S.A.}
\newcommand{\Portsmouth}{University of Portsmouth, Portsmouth, PO1 3FX, United Kingdom}
\newcommand{\BHAM}{School of Physics and Astronomy and Institute for Gravitational Wave Astronomy, University of Birmingham, Edgbaston, Birmingham, B15 2TT, United Kingdom
}

\definecolor{darkgreen}{RGB}{0,153,0}
\definecolor{epurple}{RGB}{216,131,183}

\usepackage[acronym,shortcuts]{glossaries}

\newacronym{bbh}{BBH}{binary black hole}

\begin{document}


\def\phenom{\textsc{Phenom}}
\def\seob{\textsc{SEOBNR}}
\def\nrsur{\textsc{NRSurrogate}}

\def\xcp{\textsc{IMRPhenomXHM-CP~(XHM-CP)}\gdef\xcp{\textsc{XHM-CP}}}

\def\vFivePHM{\textsc{SEOBNRv5PHM~(SEOBv5)}\gdef\vFivePHM{\textsc{SEOBv5}}}
\def\vFourPHM{\textsc{SEOBNRv4PHM}} 

\def\nrsurSDQF{\textsc{NRSur7dq4~(NRSur)}\gdef\nrsurSDQF{\textsc{NRSur}}}

\def\pd{\textsc{PhenomD}} 
\def\pvtwo{\textsc{PhenomPv2}} 
\def\phm{\textsc{PhenomHM}} 
\def\pvthree{\textsc{PhenomPv3}} 
\def\pvthreeHM{\textsc{PhenomPv3HM}} 
\def\pnr{\textsc{PhenomPNR~(PNR)}\gdef\pnr{\textsc{PNR}}}
\def\pdcp{\textsc{PhenomDCP~(DCP)}\gdef\pdcp{\textsc{DCP}}}
\def\px{\textsc{IMRPhenomX}} 
\def\pxas{\textsc{IMRPhenomXAS~(XAS)}\gdef\pxas{\textsc{XAS}}}
\def\pxhm{\textsc{IMRPhenomXHM}} 
\def\pxphm{\textsc{IMRPhenomXPHM~(XPHM)}\gdef\pxphm{\textsc{XPHM}}}
\def\spintaylor{\textsc{XPHM-ST}}
\def\tphm{\textsc{IMRPhenomTPHM~(TPHM)}\gdef\tphm{\textsc{TPHM}}}
\def\modelname{\textsc{PhenomXO4}a~\textsc{(XO4}a)\gdef\modelname{\textsc{XO4}a}}

\def\modelnameLAL{\texttt{IMRPhenomXO4a}}

\def\gw#1{gravitational wave#1}
\def\nr#1{numerical relativity
 (NR)#1\gdef\nr{NR}}
\def\bh#1{black-hole
 (BH)#1\gdef\bh{BH}}
\def\bbh#1{binary black hole#1
  (BBH#1)\gdef\bbh{BBH}}
\def\qnm#1{Quasinormal Mode
    (QNM)#1\gdef\qnm{QNM}}
\def\pn#1{post-Newtonian (PN)#1\gdef\pn{PN}}
\def\imr#1{inspiral-merger-ringdown (IMR)#1\gdef\imr{IMR}}
\def\eob#1{effective-one-body
   (EOB)#1\gdef\eob{EOB}}
\def\oed#1{optimal emission direction#1}
\def\pe#1{parameter estimation (PE)#1\gdef\pe{PE}}

\def\mism{\mathfrak{M}}
\def\pmism{\mathfrak{M}_\text{w}}
\def\Eqn#1{Equation~(\ref{#1})}
\def\eqn#1{Eq.~(\ref{#1})}
\def\ceqn#1{Eq.~\ref{#1}}
\newcommand{\Eqns}[2]{Equations~(\ref{#1}-\ref{#2})}
\newcommand{\Eqnsa}[2]{Equations~(\ref{#1}) and (\ref{#2})}
\newcommand{\eqns}[2]{Eqs.~(\ref{#1}-\ref{#2})}
\newcommand{\eqnsa}[2]{Eqs.~(\ref{#1}) and (\ref{#2})}
\newcommand{\ceqns}[2]{Eqs.~\ref{#1}-\ref{#2}}
\newcommand{\ceqnsa}[2]{Eqs.~\ref{#1} and \ref{#2}}

\def\PaperOne{\hyperlink{cite.Hamilton:2021pkf}{Paper~{I}} }

\title{\textsc{PhenomXO4}a: a phenomenological gravitational-wave model for precessing
black-hole binaries \\ with higher multipoles and asymmetries}


\author{Jonathan E. Thompson}
\affiliation{\Caltech}
\affiliation{\Cardiff}

\author{Eleanor Hamilton}
\affiliation{\Zurich}
\affiliation{\UIB}

\author{Lionel London}
\affiliation{\Amsterdam}
\affiliation{\KCL}

\author{Shrobana Ghosh}
\affiliation{\AEI}
\affiliation{\Leibniz}
\affiliation{\Cardiff}

\author{Panagiota Kolitsidou}
\affiliation{\Cardiff}
\affiliation{\BHAM}

\author{Charlie Hoy}
\affiliation{\Portsmouth}

\author{Mark Hannam}
\affiliation{\Cardiff}

\begin{abstract}
In this work we introduce \textsc{PhenomXO4}a, the first
phenomenological, frequency-domain gravitational waveform model
to incorporate multipole asymmetries
and precession angles tuned to numerical relativity.
We build upon the modeling work that produced the \textsc{PhenomPNR} model and incorporate
our additions into the \textsc{IMRPhenomX} framework, retuning the coprecessing frame
model and extending the tuned precession angles to higher signal multipoles. 
We also include, for the first time in frequency-domain models, 
a recent model for spin-precession-induced
multipolar asymmetry in the coprecessing frame to the dominant gravitational-wave multipoles.
The accuracy of the full model and its constituent components is assessed through comparison 
to numerical relativity and numerical relativity surrogate waveforms by computing mismatches and 
performing parameter estimation studies. We show that, for the dominant signal multipole, we retain the 
modeling improvements seen in the \textsc{PhenomPNR} model. We find that the relative accuracy 
of current full IMR models varies depending on location in parameter space and the comparison 
metric, and on average they are of comparable accuracy. However, we find that variations
in the pointwise accuracy do not necessarily translate into large biases 
in the parameter estimation recoveries.
\end{abstract}

\date{\today}

\maketitle

\section{Introduction}

The properties of compact-binary gravitational-wave (GW) sources are inferred by convolving detector data with
theoretical signal models~\cite{KAGRA:2013rdx,LIGOScientific:2014pky,VIRGO:2014yos,KAGRA:2020tym,LIGOScientific:2018mvr,LIGOScientific:2020ibl,LIGOScientific:2021usb,LIGOScientific:2021djp, Nitz:2021zwj, Venumadhav:2019lyq, Mehta:2023zlk}. In LIGO-Virgo-KAGRA (LVK) compact binary observations to date the most commonly used families of
signal models have been \phenom{}, \seob{} and \nrsur{}~\cite{Husa:2015iqa, Khan:2015jqa, Hannam:2013oca, London:2017bcn, Dietrich:2019kaq, Khan:2019kot, Thompson:2020nei, Pratten:2020fqn, Garcia-Quiros:2020qpx, Pratten:2020ceb, Estelles:2021gvs, Taracchini:2013rva, Pan:2013rra, Cotesta:2018fcv, Ossokine:2020kjp, Matas:2020wab, Pompili:2023tna, Ramos-Buades:2023ehm, Blackman:2017dfb, Varma:2018mmi, Varma:2019csw,Islam:2021mha} 
The \phenom{} and \seob{} models are constructed from a combination of analytic or semianalytic
approximations during the inspiral, and \nr{} calculations for the late inspiral, merger
and ringdown; the \nrsur{} models are constructed primarily from \nr{} waveforms. 
The accuracy of \phenom{}
and \seob{} models is determined by the accuracy of the analytic ingredients, by the length, accuracy and
parameter-space coverage of the \nr{} waveforms, and by the details of the model
construction, including any physical approximations. By contrast, the main limitation in current surrogate models 
is not their accuracy, but their parameter-space coverage and the length of the input waveforms
(which, given the fixed frequency range of detector sensitivities, limits the masses 
at which they can be used for analysis).

Two of the limitations in the \phenom{} and \seob{} models used in the first three LVK observing runs (O1-3) were
that precession effects (due to spins misaligned with the orbital angular momentum) are based only on analytic
approximations and do not include tuning to any \nr{} simulations of precessing binaries, and that the models neglect
an asymmetry in the multipoles that is present in precessing configurations. In this paper we present a new model that
adds both of these features, \modelname{}. (Throughout the paper we will introduce each model with its full name, but
also introduce an abbreviation, which we we will use to simplify reference throughout the paper to a large number of 
models with very similar names.)

This paper extends on the work in Ref.~\cite{Hamilton:2021pkf} (\PaperOne\!\!), and we refer the reader to this paper for
more background on the phenomenology of precessing binaries, and a summary of approaches to modelling the GW 
signal from them. \PaperOne presented the \pnr{} model, where merger-ringdown precession effects in the $\ell=2$ 
multipoles were tuned to 40 \nr{} simulations of single-spin systems between mass ratios of $1 \le q = m_1/m_2 \le 8$, 
where \(m_1\) and \(m_2\) are the component masses of the binary system with \(m_1\ge m_2\), and the primary black 
hole has dimensionless spin $\chi$ misaligned from the orbital angular momentum by $\theta_{\rm LS}$. 

In this work the model is retuned to all 80 simulations in the \nr{} catalogue discussed in Ref.~\cite{Hamilton:2023qkv},
to improve the overall accuracy of the phenomenological fits, but in particular
the behavior of low-spin binaries near the aligned-spin limit. The resulting model is implemented within the 
\px{} infrastructure~\cite{Pratten:2020ceb}.
We also extend the model of the precession dynamics to higher multipoles through an approximate frequency mapping
analogous to that used to construct higher signal multipoles in the earlier \phm{} model~\cite{London:2017bcn}, 
and make use of an estimate of the ringdown frequency for each multipole in a frame that tracks the binary's 
precession, as presented in Ref.~\cite{Hamilton:2023znn}.

We incorporate a model for the multipole asymmetry in the dominant \((\ell,|m|)=(2,2)\) multipoles based on a prescription
presented in Ref.~\cite{Ghosh:2023mhc}. This allows us to easily construct the antisymmetric contribution to the
dominant multipole from physical quantities that were already modeled for the symmetric contribution, namely
the signal amplitude, phase, and one of the three Euler angles that specify the precession.

The current model is the first frequency-domain higher-multipole inspiral-merger-ringdown model to include precession tuning to \nr{}
and multipole asymmetry, and provides a first indication of the accuracy improvements each new feature
provides, and improvements that need to be made in future to meet the accuracy needs of gravitational-wave
observatories.

One issue that arises in frequency-domain models is with the now-standard procedure to separately model 
(a) the signal in a coprecessing frame that tracks the precession, and (b) the time- or frequency-dependent 
rotation transformation between the inertial and co-precessing frames. These two models are then combined 
to produce the final inertial-frame model. This approach is motivated by the observation that the signal multipoles
take a simpler form in the co-precessing frame~\cite{Schmidt:2012rh}, and is used in some form in all precessing 
\phenom{}, \seob{} and \nrsur{} models.

In the time domain the co-precessing frame defined with respect to the $\ell=2$ multipoles is approximately the 
same as that defined with respect to all multipoles, i.e., the directions that maximise the power emitted by either the 
$\ell=2$ multipoles or all multipoles are approximately the same. This is not the case in the frequency domain. 
This is easiest to see when considering the ringdown. The ringdown frequency of the $(\ell,|m|)=(4,4)$ multipole
is roughly twice that of the $(\ell, |m|)=(2,2)$ multipole, and so in the frequency domain the $(4,4)$ ringdown
will begin at a frequency where there is no longer any power remaining in the $(2,2)$. By contrast, in the time domain 
the ringdown begins at roughly the same time for all multipoles. We discuss this issue further in 
Sec.~\ref{sec:frame-choices}, and the approximations we use to circumvent the issue in the current model.

A number of earlier models are referred to throughout this paper. For ease of reference we summarize them 
here. We start with \phenom{} models. The aligned-spin \phenom{} model \pd{}~\cite{Husa:2015iqa,Khan:2015jqa} 
includes only the dominant multipoles, and was tuned to simulations of (predominantly) single-spin or equal-spin 
binaries, up to mass ratios of $q=18$. This was used as the basis of the coprecessing-frame model for 
the precessing-binary models \pvtwo{}~\cite{Hannam:2013oca} and \pvthree{}~\cite{Khan:2018fmp}; \nr{} calibration to 
in-plane-spin modifications are added to produce \pdcp{}, which is the co-precessing-frame model for the 
\nr{}-tuned precession model \pnr{}~\cite{Hamilton:2021pkf}. The more recent aligned-spin models for the 
dominant multipoles (\pxas{}~\cite{Pratten:2020fqn}) and higher multipoles (\pxhm{}~\cite{Garcia-Quiros:2020qpx}) 
are tuned to NR simulations of unequal-spin binaries, and this is the basis of the coprecessing-frame model for the 
precessing model \pxphm{}~\cite{Pratten:2020ceb}. In this work we incorporate \nr{} in-plane-spin tuning to
produce \xcp{}.

The precession dynamics are modeled in \pvthree{} using a multi-scale analysis (MSA) approach~\cite{2017PhRvL.118e1101C},
and these are also adopted in \pxphm{}. The MSA dynamics are augmented by \nr{}-tuned merger-ringdown modeling 
in \pnr{}, which we extend in this work. Alternatively, one may use a frequency-domain parameterization of the
time-domain spin evolution used in the time-domain Phenom model \tphm{}~\cite{Estelles:2021gvs}; these angles
are used in the \spintaylor{} model~\cite{Colleoni:2023}, and an efficient method for solving and implementing 
them is found in \textsc{IMRPhenomXODE}~\cite{Yu:2023lml}.

The \seob{} models integrate the EOB equations of motion (with additional \nr{} tuning) to calculate both the inspiral 
phasing and precession dynamics, followed by smoothly connecting ringdown modes at merger. These models are
tuned only to aligned-spin \nr{} waveforms, although precessing-binary waveforms are used for verification of the 
final model. In this paper we compare against the most recent \seob{} model, \vFivePHM{}~\cite{Ramos-Buades:2023ehm}.

Finally, the \nrsur{} model \nrsurSDQF{}~\cite{Varma:2019csw} is calibrated to precessing \nr{} waveforms up to $q=4$, 
and can be used to analyze signals starting around \(20\)~Hz with masses $M \gtrsim 65\,M_\odot$. Comparisons against \nr{} waveforms suggest
that this is the most accurate model currently available, and we use it for comparisons where possible. However, since it 
cannot be used for low-mass signals, and its calibration region does not extend beyond $q=4$ (we calibrate our model
up to $q=8$), it cannot provide a definitive test of our model across the full calibration parameter space. (Conversely,
the lack of an extremely accurate full inspiral-merger-ringdown model across the full binary parameter space is the 
primary motivation for the work we present here.)

Sec.~\ref{sec:cp-all} outlines how the original \pdcp{} tuning is improved 
and applied to \pxas{}~\cite{Pratten:2020fqn}, 
as well as introduces the model for the antisymmetric 
contributions added to the coprecessing \((\ell,|m|)=(2,2)\) multipoles. 
Information about the precession 
angles, including modifications to the \pnr{} angles and the 
mapping of these angles to higher multipoles is found 
in Sec.~\ref{sec:angles}. The performance of the final model is 
considered in Sec.~\ref{sec:model-performance},
presenting mismatch and parameter estimation results compared with 
other contemporary waveform approximants, and we 
end with concluding remarks in Sec.~\ref{sec:conclusions}.

In this paper we use geometric units, whereby \(G=c=1\). Unless otherwise stated, all frequencies
are presented in dimensionless units \(Mf\). We employ use of the symmetric mass-ratio 
\(\eta = m_1 m_2 / (m_1+m_2)^2\), the effective aligned-spin parameter 
\(\chi_\text{eff}\)~\cite{Ajith:2009bn,Santamaria:2010yb} and the 
precession spin parameter \(\chi_\text{p}\)~\cite{Schmidt:2014iyl}.
Finally we remark that results in this paper were produced using the reviewed implementation
of this model in \texttt{LALSuite}~\cite{lalsuite,swiglal}, where 
the model is called \modelnameLAL.

\section{On the importance of frame choices}
\label{sec:frame-choices}

Precession introduces modulations in the amplitude and phase of a gravitational-wave signal, and this complicates both the
signal and, as a result, the task of modeling it. One common modeling technique is to make a time-dependent rotation to a
``coprecessing'' frame that tracks the orbital precession. In such a frame the constituent parts of the signal --- the coprecessing-frame
waveform and the time-dependent rotation angles (with respect to the total angular momentum, $\mathbf{J}$) --- take on simple forms that make them 
easier to model.
In addition, during the inspiral the coprecessing-frame signal can be approximated well by the signal from an equivalent non-precessing
binary~\cite{Schmidt:2012rh}.
This is possible because, to a good approximation, the orbital frequency and inspiral rate are determined by the black-hole
masses and the spin components that are aligned with the orbital angular momentum, i.e., the spins that constitute an aligned-spin system,
and this orbital motion is the dominant contribution to the GW signal; the precession due to the in-plane spin components
can be considered as a time-dependent rotation applied to this aligned-spin waveform.

This simplification, which is easy to motivate in the quadrupole approximation to the signal
from an orbiting binary, also extends to the sub-dominant multipoles~\cite{Schmidt:2010it}, and, although the aligned-spin mapping is
no longer valid through merger and ringdown~\cite{Schmidt:2012rh,Pekowsky:2013ska,Ramos-Buades:2020noq}, 
the coprecessing-frame waveform and the time-dependent precession angles retain a sufficiently simple morphology that the 
coprecessing frame continues to be attractive for signal modeling.

Some form of coprecessing-frame decomposition has been used in all precessing-binary models that have been used to analyze
LVK observations to date~\cite{Hannam:2013oca, Khan:2019kot, Pratten:2020ceb, Estelles:2021gvs, Hamilton:2021pkf, Taracchini:2013rva, Pan:2013rra, Ossokine:2020kjp, Ramos-Buades:2023ehm, Varma:2019csw}.
The coprecessing frame can be defined either using the orbital dynamics of the binary,
or a frame aligned with the maximum strength
of the gravitational-wave signal, referred to as either quadrupole alignment~\cite{Schmidt:2010it} or the optimal emission
direction~\cite{OShaughnessy:2011pmr,Boyle:2011gg}. The \seob{}~\cite{Pan:2013rra,Taracchini:2013rva,Ossokine:2020kjp} and 
\phenom{}~\cite{Hannam:2013oca, Khan:2019kot, Pratten:2020ceb, Estelles:2021gvs,Yu:2023lml} models used the precession defined via the orbital dynamics, while the more recent \pnr \ model~\cite{Hamilton:2021pkf},
and the NRSurrogate models~\cite{Blackman:2017dfb,Varma:2019csw} use the optimal emission direction, which is the correct
transformation between the coprecessing frame and the inertial frame of the \gw~signal; the differences between coprecessing
frames defined with respect to either the \emph{signal} or the \emph{dynamics}, are discussed further in 
\\ \PaperOne\!.

Unfortunately, the approximate mapping between precessing- and aligned-spin-binary waveforms does not carry over to the frequency
domain beyond the dominant coprecessing-frame multipoles, $(\ell,|m|)=(2,2)$: the optimal emission direction no longer
identifies a frame in which the signal multipoles can be approximated by their aligned-spin counterparts.

To illustrate this, first consider the early inspiral in a coprecessing frame in the time domain. We make use of the standard
decomposition of the \gw~strain, $h(t)$, into spin-weighted spherical harmonics, 
\begin{equation}
h(t) = \sum_{\ell, m} h_{\ell,m}(t) \, ^{-2}Y_{\ell,m}(\theta,\phi),
\end{equation} where $(\theta,\phi)$ are the standard polar and azimuthal angles in spherical polar coordinates, and 
$\ell\geq2$ and $|m|\leq \ell$. 
As already noted, in this coprecessing frame
the signal multipoles will approximate those of an aligned-spin, non-precessing binary, i.e., \begin{equation}
h^{\rm{NP}}_{\ell m}(t) = A_{\ell m}(t) e^{- i m \Phi(t)},   \label{eq:CPmultipoles}
\end{equation} where $\Phi(t)$ is the orbital phase of the binary and $A_{\ell m}(t)$ the amplitude.
At each time~$t$, we can rotate the multipoles using a set
of Euler angles $\left(\alpha(t), \beta(t), \gamma(t)\right)$ to produce the multipoles in the inertial frame.

Now consider the same early inspiral signal in the frequency domain. We see from Eq.~(\ref{eq:CPmultipoles}) that the frequency of
each multipole scales with $m$. 
If at time $t_0$ the frequency of the dominant $(\ell,|m|)=(2,2)$ multipole is $f_0$, then the frequency
of each other multipole is $m f_0 / 2$. This means that the angles $(\alpha(t), \beta(t), \gamma(t))$ should be applied to the (2,2)
multipoles at $f_0$, but to the (3,3) multipoles at $3f_0/2$, the (4,4) multipoles at $2 f_0 $, and so on. Conversely, if we now consider
the signal at only one frequency, $f_0$, then the angles $(\alpha, \beta, \gamma)$ applicable to rotate back to the inertial frame will be different
for each $m$. (This is what is currently done in the stationary phase approximation
[SPA] treatment of the twisting-up procedure in the \pvthreeHM{} and \pxphm{}
models~\cite{Pratten:2020ceb,Khan:2019kot}.)

Therefore, we cannot rotate between the coprecessing frame as defined above 
and the inertial frame at each frequency by a single set of Euler angles. 
Since the optimal emission direction \emph{would} be defined
by a single rotation of all of the multipoles at each frequency, it cannot be identified with the Fourier transform of the time-domain 
coprecessing frame waveform,
unless we restrict our coprecessing-frame signal to one value of $m$. This restriction is made in the dominant-multipole frequency-domain
models \pvtwo{} and \pvthree{}, but not in the higher-multipole models \pvthreeHM{} or \pxphm.

If we wish to produce a frequency-domain model that comprises a coprecessing-frame model 
plus a model for the precession angles, then
we must either define a new coprecessing frame that includes all $\ell$ and reconsiders the current aligned-spin mapping, 
or identify a generalization of the SPA frequency mapping that is valid through the merger and ringdown. In the model we present here, 
we choose the latter. For the
remainder of this paper we address the problem of how to define and model a frequency-domain 
coprecessing-frame signal that retains
the aligned-spin mapping during the inspiral.

\section{Coprecessing Frame}
\label{sec:cp-all}
In this section, we describe the symmetric and coprecessing sector of \modelname{}, which includes NR-tuned modifications that encode precession effects.
These modifications are directly built upon \pxhm{}~\cite{Garcia-Quiros:2020qpx}, which is a model for \gw{} signals from non-precessing \bbh{s}, and 
the basis of the coprecessing waveform for \pxphm{}.
Our approach for modifying \pxhm{} mirrors that used for \pdcp{}, which is the coprecessing model presented in \PaperOne\!.
\par In Sec.~\ref{sec:cp} we describe both the tuning of the dominant symmetric \((\ell,|m|)=(2,2)\) coprecessing multipoles, extending the work that produced \pdcp{} in \PaperOne\!, and the use of an ``effective ringdown frequency''~\cite{Hamilton:2023znn} in the higher coprecessing signal multipoles.
The addition of antisymmetric contributions to the \((\ell,|m|)=(2,2)\) coprecessing multipoles, modeled in Ref.~\cite{Ghosh:2023mhc}, is detailed in Sec.~\ref{sec:asymmetry}.

\subsection{Higher order multipole coprecessing model}
\label{sec:cp}

\subsubsection{$\ell=m=2$ multipole calibration}
\label{sec:cp-22}
We calibrate the $(2,2)$ multipole moments to \nr{} simulations.
Our approach follows that of \PaperOne\!\!, in that calibration is done by applying deviations to \pxas{} model parameters. \pxas{} is the 
successor to \pd{} within the \px{} framework and describes the \((\ell,|m|)=(2,2)\)
emission in the \pxhm{} model.  
If $\lambda_k$ is an \pxas{} model parameter, then our coprecessing model uses the modified parameter $\lambda'_k$,
\begin{align}
   \label{dev1}
    \lambda_k' \;  = \; \lambda_k \, + \, \chi  \sin(\theta_{\mathrm{LS}}) \, u_k  ,
\end{align}
where $u_k$ is a \textit{deviation variable} that is tuned to \nr{}, and $\chi$ and $\theta_{\rm LS}$ are the primary black hole's
spin magnitude and misalignment (as in the single-spin simulations used to tune the model), or, equivalently, the
result of the single-spin mapping of generic two-spin binaries, as described in \PaperOne\!.
In implementation, $u_k$ is denoted differently, according to physical significance:
\begin{align}
   \label{eq:deviation variables}
    u_k \in \{\mu_1,\mu_2,\mu_3,\nu_0,\nu_4,\nu_5,\nu_6,\zeta_1,\zeta_2\}\;.
\end{align}
The physical significance of each version of $u_k$ is noted in Table~\ref{tab:cp-pars}.
\par Numerical tuning is achieved by first parameterizing \pxas{} by the set of $n$ deviation variables $\{u_0,u_1,...u_n\}$ using the \texttt{LALDict} infrastructure.
This results in a version of \pxas{} that is deformable ``on-the-fly", meaning that \texttt{LALSuite}'s \texttt{swig-python} interface allows for \texttt{Python} generation of waveforms that have amplitudes and phase derivatives determined by the choice and values of each $u_k$.
As with \pdcp{}~\cite{Hamilton:2021pkf}, the deviation variables are applied to select amplitude and phase model parameters.
\par On-the-fly tuning of select amplitude and phase parameters results in a new model for the coprecessing frame multipole moments that we will call \xcp{}.
\par The tuning of each $u_k$ entails (in order) determining a minimal set of model parameters to tune, numerically solving for optimal values of $u_k$ for each calibration case (i.e., tuning), modeling the set of optimal deviation variables across $(\chi,\theta_{\mathrm{LS}},\eta,\delta=\sqrt{1-4\eta})$, and then allowing the on-the-fly deformable version of \pxas{} to be generated with modeled optimal deviation variables. 
\par The minimal set of deviation variables is listed in Table~\ref{tab:cp-pars}.
The parameters listed there were selected to enable modifications of \pxas{} to fit the \nr{} waveforms (amplitude and phase derivatives) within our calibration set.
Given the choice of deviation parameter, and this set of deviation variables, optimal values of the deviation variables were found by minimizing a representation error, which we defined to be the sum of two positive definite quantities: the root-mean-square error (RMSE) for the frequency domain waveform's amplitude, and the RMSE for the frequency domain phase derivative.
Minimization of representation error was performed using \texttt{scipy.optimize}.
\par Each deviation variable was modeled as a multivariate polynomial using a significantly refined version of the basis learning routine, \texttt{gmvpfit}, detailed in Ref.~\cite{London:2018nxs}.
To avoid over-fitting, the minimum allowed fractional change in representation error, i.e. \texttt{gmvpfit}'s \texttt{estatol} keyword input, was set to $0.001$.
Further information about the specific tuning coefficients is found in Appendix~\ref{sec:waveform-flags}.
\par Outside of the calibration region (i.e. the runs presented in Ref.~\cite{Hamilton:2023qkv}; $q>8$, $\chi>0.8$, $\theta_{LS}>150^\circ$), \xcp{} transitions smoothly to a version of \pxhm{} that is modified with the effective ringdown frequencies for precessing \bbh{} remnants~\cite{Hamilton:2023znn}.
This transition occurs separately along parameter space coordinates $q$, $\chi$, and $\theta_{LS}$ according to a shifted $\cos$ taper,
\begin{align}
   \label{pnr-window}
   w_v \; = \; \frac{1}{2}\left[ \cos\left( \frac{v-v_b}{v_w} \right) + 1 \right] \; .
\end{align}
In \eqn{pnr-window}, $v$ is one of the coordinate directions, i.e. $v\in\{q,\chi,\theta_{LS}\}$, $v_b$ is a calibration boundary, and $v_w$ describes the transition width.
For $\{q,\chi,\theta_{LS}\}$, the respective values of $v_b$ are $\{8,0.8,150^\circ\}$.
Note that while the factor of $\sin(\theta_\mathrm{LS})$ in \eqn{dev1} ensured that parameter deviations are zero in the spin-aligned and anti-aligned limits, use of windowing in the $\theta_\mathrm{LS}$ direction has been found to slightly improve the smoothness of parameter deviations between $\theta_{LS}=150^\circ$ and $\theta_{LS}=180^\circ$.
For $\{q,\chi,\theta_{LS}\}$, the respective values of $v_w$ are $\{0.5,0.02,0.5\}$.
\par On the tuned parameter space's boundaries, \eqn{dev1} takes the form
\begin{align}
   \label{dev2}
    \lambda_k' \;  = \; \lambda_k \, + 
    \, W(q,\chi,\theta_\text{LS})
    \; \chi  \sin(\theta_{\mathrm{LS}}) \, u_k  \; ,
\end{align}
where
\begin{align}
   W(q,\chi,\theta_\text{LS}) \; = \;  w_q 
   \, w_{\theta_{LS}} 
   \, w_{\chi}  \; .
\end{align}
Concurrently, since the \xcp{} tuning is applied directly atop \pxhm{} (of which \pxas{} is a component), the remnant's spin must transition from the non-precessing spin within our calibration region, to the appropriate precessing final spin outside.
Thus on the parameter space boundaries, the final \bh{}'s dimentionless spin, $a_f$, takes on the following form 
\begin{align}
   a_f \; = \;  (1-W)a^\text{Prec.}_f + W a^\text{Non-Prec.}_f \; ,
\end{align} 
as is needed to self-consistently determine related \pxhm{} model parameters, and the ringdown frequency appropriate for precessing systems~\cite{Garcia-Quiros:2020qpx,Hamilton:2023znn}.

\subsubsection{Subdominant multipole moments}
\label{sec:hm-coprec}

We do not directly calibrate the $(\ell,|m|)\neq (2,2)$ multipole moments. Instead, we adjust only the ringdown frequency of the multipoles to account for the fact that the waveform is modeled in the coprecessing frame.  This frame is different to that in which the ringdown frequencies are calculated from perturbation theory, where the final spin is along the $z$-direction. The correct treatment of the ringdown frequency for precessing systems, and the relationship between the co-precessing and inertial frames, is discussed in greater detail in Ref.~\cite{Hamilton:2023znn}.

The effective ringdown frequency in the coprecessing frame is given by~\cite{Hamilton:2023znn},
\begin{align} \label{eqn: HOM expression}
   \omega'_{\ell m} =
   {}& \omega_{\ell m} - m(1-|\cos\beta_\mathrm{f}|)\left(\omega_{22} - \omega_{21}\right).
\end{align}
where $\beta_\mathrm{f}$ is the final ringdown value of $\beta$ taken from the fits discussed in Sec.~\ref{sec:dominant-angles} and $\omega_{\ell m}$ are the ringdown frequencies calculated from perturbation theory. In order to obtain the mass and spin of the final black hole required to obtain the ringdown frequencies, we use the aligned-spin fit for the final mass given in~\cite{Jimenez-Forteza:2016oae} and a final spin given by
\begin{align}
   \boldsymbol{\chi}_\mathrm{f} = {}& \mathrm{sgn}\left(\cos\beta_\mathrm{f}\right)\chi_\mathrm{f},
\end{align}
where $\chi_\mathrm{f}$ is given by Eq.~(\ref{eqn: final spin magnitude}). Taking the sign of the spin to correspond to the sign of $\cos\beta_\mathrm{f}$ ensures that the prograde or retrograde frequency is chosen correctly as we move across the parameter space. The conceptual framework and physical motivation behind this choice is described in detail in Sec. IV. A of Ref.~\cite{Hamilton:2023znn}.

\subsection{Multipole asymmetry}
\label{sec:asymmetry}

Current inspiral-merger-ringdown waveform approximants (from both the \phenom{} and \seob{} families) model the $-m$ coprecessing-frame multipoles
using the reflection symmetry,
\begin{equation}
h^{\rm CP}_{\ell m} = (-1)^{\ell} h^{{\rm CP}\, *}_{\ell-m}\;.
\end{equation}
This symmetry holds for aligned-spin binaries, but is broken for systems with misaligned spins~\cite{Arun:2008kb,Boyle:2014ioa,Kalaghatgi:2021inv,Ramos-Buades:2020noq}.
This asymmetry results in
linear momentum radiation perpendicular to the orbital plane, which can lead to large out-of-plane recoil in the final
black hole~\cite{Bruegmann:2007bri}. The phasing of the antisymmetric signal contribution is affected by the direction of the in-plane spin components,
and for this reason we expect the inclusion of the antisymmetric contribution in signal models to improve the accuracy of
the black-hole-spin measurements in GW observations. The NR surrogate model \nrsurSDQF{}~\cite{Varma:2019csw} \emph{does}
include the antisymmetric contribution,
and it was shown in Ref.~\cite{Kolitsidou:2023} that this contribution is indeed necessary to measure the full spin information,
 and was also likely crucial in identifying precession in the analysis of GW200129 in Ref.~\cite{Hannam:2021pit}; we expect the
 same to be true for the measurement of  a large recoil in GW200129 in Ref.~\cite{Varma:2022pld}.

Ref.~\cite{Ghosh:2023mhc} introduces a method to model the antisymmetric contribution to the coprecessing-frame dominant multipoles
in the frequency domain. Here we adopt this procedure and the model of the antisymmetric amplitude ratio in \cite{Ghosh:2023mhc} to construct the antisymmetric waveform from the already-existing model of the symmetric (2,2) coprecessing-frame multipoles. The \modelname{} model includes symmetric contributions to multipoles
up to $\ell=4$. As shown in Fig.~1 of Ref.~\cite{Ghosh:2023mhc}, if we neglect symmetric contributions for $\ell \geq 5$, we can also neglect higher-order
antisymmetric contributions. As such, the only antisymmetric contribution we include in \modelname{} is to the $(2,\pm 2)$ multipoles.

Specifically, the coprecessing-frame $(\ell,|m|)=(2,2)$ multipoles are split into symmetric and antisymmetric parts, \begin{eqnarray}
h_{22}(f) & = &  A_s(f) e^{i \phi_s(f)} + A_a(f) e^{i \phi_a(f)},  \\
h_{2-2}(f) & = & A_s(f) e^{-i \phi_s(f)} - A_a(f) e^{-i \phi_a(f)}.
\end{eqnarray}
The symmetric (2,2) amplitude and phase, $A_s(f)$ and $\phi_s(f)$, are the amplitude and phase of the
standard model that does not include multipole asymmetries. The antisymmetric amplitude $A_a(f)$ is constructed as a
rescaling of the symmetric amplitude, so
\begin{equation}
A_a(f) = \kappa(f) A_s(f),
\end{equation}
where the model of the ratio $\kappa(f)$ is made up of a \pn{} estimate and an
NR-calibrated correction, described by a single coefficient $b$ as defined in Eq.~(16) in Ref.~\cite{Ghosh:2023mhc}. Note that the coefficient $b$ showed no strong correlation with spin magnitude, while the spin dependence of the amplitude ratio was carried over from the \pn{} estimate. Therefore, $\kappa(f)$ was calibrated to the 80 simulations presented in Ref.~\cite{Hamilton:2023qkv} by fitting an ansatz for $\kappa$ across the parameter space of mass-ratio and spin-misalignment. The model for the fit coefficients $b$ can be found in Eq.~(18) in Ref.~\cite{Ghosh:2023mhc}.

For single-spin systems the antisymmetric phase $\phi_a$ varies as the combination of the orbital phase and the precession angle, $\alpha$, during the inspiral, and in the
ringdown it matches the symmetric phase, i.e.,
\begin{equation} \phi_a(f) \vcentcolon =\begin{cases}
   \frac{\phi_s(f)}{2} + \alpha(f) & f< pf_{RD},\\
   \phi_s(f) & f\geq pf_{RD} \, ;
\label{asymmetryphase}
\end{cases}
\end{equation}
$f_{RD}$ is the ringdown frequency and $p \in [0,1]$ determines the fraction of $f_{RD}$ at which the transition occurs. The transition is made smooth by using a window function. The phase construction in
Eq.~(\ref{asymmetryphase}) shows an interesting unanimity across the parameter space. In particular, it was possible to use the same parameters of the window function across the full binary parameter space. Details of antisymmetric waveform calibration and full description of both the amplitude and phase models can be found in Ref.~\cite{Ghosh:2023mhc}. This model can be used to generate the antisymmetric waveform for two-spin systems by mapping it to an equivalent single-spin configuration, 
like the one used for the precession angles in~\PaperOne\!. Note that the multipole asymmetry vanishes for equal mass binaries when both spins are equal in magnitude and point in the same direction. To accommodate this behavior we modify the definition of the in-plane spin of the equivalent system, as given by Eq.~(20) in Ref.~\cite{Ghosh:2023mhc}.

To construct the positive-frequency precessing-signal strain in the frequency domain, \pxphm{} follows the prescription outlined in Appendix~E of Ref.~\cite{Pratten:2020ceb}. Noting that in the coprecessing frame,
\begin{align*}
h_{\ell m}(f) &= h^{(S)}_{\ell m}(f) + h^{(a)}_{\ell m}(f), \\
h_{\ell -m}(-f) &= h^{(S)}_{\ell -m}(-f) - h^{*(a)}_{\ell -m}(-f) \, ,
\end{align*}
it is easy to see from Eqs.~(E8) and (E9) that the symmetric and the antisymmetric contributions to the polarizations in the inertial $\mathbf{J}$-frame can be treated independently
as, \begin{eqnarray}
\tilde{h}^J_+(f>0) & = & \tilde{h}^{J,(S)}_+(f) + \tilde{h}^{J,(a)}_+(f), \\
\tilde{h}^J_\times(f>0) & = & \tilde{h}^{J,(S)}_\times(f) + \tilde{h}^{J,(a)}_\times(f).
\end{eqnarray}
The symmetric parts  $\tilde{h}^{J,(S)}_+(f)$ and $\tilde{h}^{J,(S)}_\times(f)$ are as given in Eqs.~(E18) and (E19) in Ref.~\cite{Pratten:2020ceb},
while for the antisymmetric contributions we have, \begin{equation}
\begin{split}
 \tilde{h}^{J,(a)}_+(f>0)  = &\\
 &\frac{1}{2} \sum^2_{m'>0} e^{- i m' \gamma} \tilde{h}^{(a)}_{2-m'}(f) \sum^{2}_{m=-2} \left(
 A^2_{m-m'} - A^{2*}_{mm'} \right), \\
 \end{split}
 \end{equation}
 \begin{equation}
 \begin{split}
 \tilde{h}^{J,(a)}_\times(f>0)  = &\\
 & \frac{i}{2} \sum^2_{m'>0} e^{- i m' \gamma} \tilde{h}^{(a)}_{2-m'}(f) \sum^{2}_{m=-2} \left(
 A^2_{m-m'} + A^{2*}_{mm'} \right),\\
 \end{split}
\end{equation} where we are considering only $\ell = 2$ for the antisymmetric contribution. The symbols $A^{\ell}_{mm'}(f)$ represent the transfer functions for rotating to the inertial frame from a coprecessing frame, as used in Ref.~\cite{{Pratten:2020ceb}}.

\section{Precession Angles}
\label{sec:angles}

In this section we describe the implementation of the precession angles
into \modelname{}. Modifications to the original \pnr{} model are
outlined in Sec.~\ref{sec:dominant-angles}, including some checks and
windowing to improve their performance across and outside the
calibration region. In Sec.~\ref{sec:final-beta-fit} we outline a new
fit for the final ringdown value of the \(\beta\) precession angle
which is used in the construction of the effective ringdown frequency
in Eq.~\eqref{eqn: HOM expression}~\cite{Hamilton:2023znn}. Further
small modifications to \pxphm{} to implement the \pnr{} angles
are discussed in Sec.~\ref{sec:additional-angle-changes}. Finally,
in Sec.~\ref{sec:hm-angles} we outline the procedure used to extend
the dominant \pnr{} angles to higher signal multipoles.

\subsection{Dominant-multipole precession angles}
\label{sec:dominant-angles}

\subsubsection{Basic form}
\label{sec:22angles-tuning}

\begin{figure*}[htbp]
   \centering
   \includegraphics[width=1.0\textwidth]{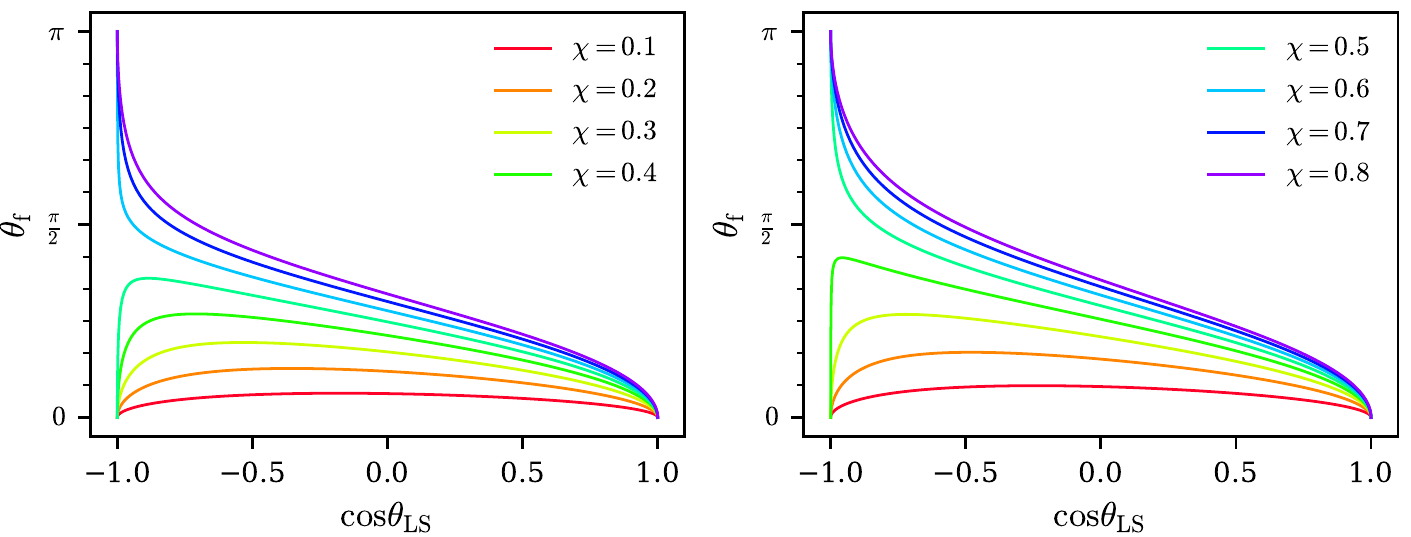}
   \caption{Evolution of the final spin inclination $\theta_\mathrm{f}$ with the inclination of the initial spin of the binary $\chi$. The left hand panel shows binaries with a mass ratio $q=6$, while the right hand panel shows those with mass ratio $q=8$. The aligned-spin limit transitions from $\theta_\mathrm{f}=0$ to $\theta_\mathrm{f}=\pi$ when $\chi\sim0.55$ for $q=6$ and $0.42$ for $q=8$-- when the aligned component of the final spin is zero, as shown in Fig.~\ref{fig: zero final spin}. This transition is now captured in the modeling of the merger-ringdown precession angle $\beta$.}
   \label{fig: thetaf behaviour}
\end{figure*}

The precession angles that describe the rotation of the dominant multipole from the coprecessing frame into the $\mathbf{J}$-frame take the same form as described in detail in \PaperOne\!.
In brief, during inspiral the precession angles $\alpha$ and $\beta$ are given by the MSA angles outlined in Ref.~\cite{Chatziioannou:2017tdw}.
The angle $\beta$ is rescaled using a PN expression so that it describes the precession of the optimal emission direction, rather than the precession of the orbital plane. 
The rescaling expression is discussed above Eq. (43) in \PaperOne\!.
\footnote{In the original paper, this expression contains an error. It is missing a factor of two in the denominator of the arctan 
argument. The correct expression is
\begin{equation}
   \beta =  2 \tan^{-1} \left( \frac{\sec(\iota/2) \left( c_0 + c_2 v^2 + c_3 v^3 \right) }{ 2 \left( d_0 + d_2 v^2 + d_3 v^3 \right) } \right). \label{eqn:betafix} \nonumber
\end{equation}
}
During merger and ringdown, the angles are given by a phenomenological ansatz fit to data from NR simulations.
The two regimes are connected as described in Sec.~VIII of \PaperOne\!.
The third precession angle $\gamma$ is then calculated numerically using the minimal rotation condition~\cite{Boyle:2011gg}.

In this iteration of the model we have recalibrated the merger-ringdown fits against the entirety of the 80 simulations presented in Ref.~\cite{Hamilton:2023qkv}.
These simulations are all single-spin binaries, where the spin is placed on the larger black hole.
We therefore use the two-spin mapping outlined in Sec.~VII~C of \PaperOne\! to obtain the form of the merger-ringdown angles for two-spin systems.

We use the same phenomenological form for the merger-ringdown angles as introduced in \PaperOne\!.
\footnote{Note the overall sign change in the expression for $\alpha$. This is to make the definition consistent with the LAL conventions.}
These are
\begin{align}
   \label{eqn: MR alpha}
   \alpha\left(f\right) - \langle \alpha\left(f\right) \rangle= {}&
   -\left(\frac{A_1}{f} + \frac{A_2\sqrt{A_3}}{A_3 + \left(f-A_4\right)^2}\right), \\
   \label{eqn: MR beta}
   \beta\left(f\right) - \langle \beta\left(f\right) \rangle = {}&
   \beta\left(f\right) - B_0 =
   \frac{B_1 + B_2 f + B_3 f^2}{1 + B_4\left(f+B_5\right)^2},
\end{align}
where the $A_i$ and $B_i$ are free coefficients. We now explicitly enforce the non-spinning and (anti-) aligned-spin limits in the expressions of these coefficients. In the case of most of these coefficients, we want the value to tend towards zero as we move into the non-precessing limit. However, $B_0$, which represents the overall amplitude of $\beta$ during ringdown, has a more complicated behavior in the non-precessing limit.

Consider an aligned-spin system. In this case, the optimal emission direction and the orbital angular momentum are aligned and both perpendicular to the plane of the binary. 
$\beta$ therefore measures the angle between the orbital angular momentum $\mathbf{L}$ and the total angular momentum $\mathbf{J} = \mathbf{L} + \mathbf{S} = \left(|\mathbf{L}| + \mathrm{sgn}\left(\mathbf{L}\cdot\mathbf{S}\right)|\mathbf{S}|\right)\mathbf{\hat{L}}$;
\begin{align}
   \cos\beta = \mathbf{\hat{L}}\cdot\mathbf{\hat{J}} = {}&
   \mathrm{sgn}\left( |\mathbf{L}| + \mathrm{sgn}\left(\mathbf{L}\cdot\mathbf{S}\right)|\mathbf{S}| \right).
\end{align}
It is therefore obvious that when the spin of the binary is aligned with the orbital angular momentum, we have $\beta=0$. However, in the case that the spin and the orbital angular momentum are anti-aligned, if $|\mathbf{L}|>|\mathbf{S}|$, then $\beta=0$ as before, but if $|\mathbf{L}|<|\mathbf{S}|$ then $\beta=\pi$.

For us to be able to apply this intuition to precessing systems, we would need to know the magnitude of the orbital angular momentum just prior to merger. Since this is a poorly defined quantity in the non-linear regime, it is not possible to directly infer the behavior of $\beta$ as it tends towards the anti-aligned-spin limit. Instead, we use the estimate of the final spin direction, relative to the direction of the orbital angular momentum.

\begin{figure}[htbp]
   \centering
   \includegraphics[width=0.48\textwidth]{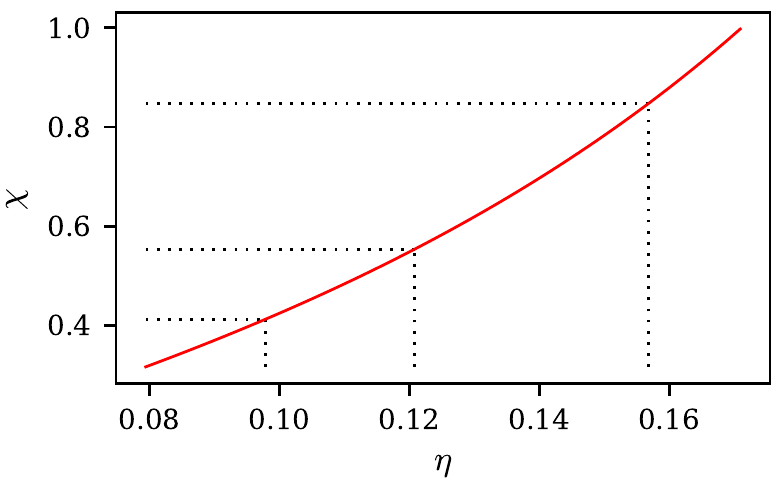}
   \caption{Variation of the value of initial binary spin $\chi$ at which the aligned component of the final spin goes to zero. Shown in dotted black are configurations with $q=8, 6, 4$, shown from left to right. Capturing the correct limiting behavior of $\beta$ is therefore particularly important to model precessing systems with higher mass ratio.}
   \label{fig: zero final spin}
\end{figure}

The final spin for a precessing binary is calculated by combining the aligned component $\chi_\mathrm{f}^\parallel$ obtained from numerical fits for aligned-spin binaries and the in-plane component $\chi_\mathrm{f}^\perp$, which is assumed to be conserved throughout the evolution of the binary. We use the aligned-spin fits given in Ref.~\cite{Jimenez-Forteza:2016oae} with the aligned-spin component given by our single-spin mapping: $\chi_1^\parallel = \chi\cos\theta_\mathrm{LS}$; $\chi_2^\parallel=0$. The inclination of the final spin relative to the orbital angular momentum just prior to merger is then given by
\begin{align}
   \cos\theta_\mathrm{f} = {}& \frac{\chi_\mathrm{f}^\parallel}{\chi_\mathrm{f}},
\end{align}
where
\begin{align} \label{eqn: final spin magnitude}
   \chi_\mathrm{f} = {}&
   \sqrt{\left(\chi_\mathrm{f}^\parallel\right)^2 + \left(\left(\frac{q}{1+q}\right)^2\chi_\mathrm{f}^\perp\right)^2}.
\end{align}
For an aligned spin binary, we find $\chi_f = |\chi_\mathrm{f}^\parallel|$ so $\cos\theta_\mathrm{f} = \pm 1$ and so recover the correct limit. We find the transition in limiting behavior occurs when $\chi_\mathrm{f}^\parallel = 0$. This is demonstrated in Figs.~\ref{fig: thetaf behaviour} and~\ref{fig: zero final spin}. The mean value of $\beta$ in the merger-ringdown regime, $B_0$ is modeled as a perturbation on top of the final spin inclination.

Having now established the correct limiting behavior of the precession angles, the coefficients in Eqs.~\eqref{eqn: MR alpha} and~\eqref{eqn: MR beta} are given by
\begin{align}
   A_1 = {}& \chi\sin\theta_\mathrm{LS}\left(\Lambda^A_1\right)^2, \\
   A_2 = {}& -\chi\sin\theta_\mathrm{LS}\left(\Lambda^A_2\right)^2, \\
   A_3= {}& \left(\Lambda^A_3\right)^2, \\
   A_4 = {}& \left(\Lambda^A_4\right)^2, \\
   \langle \beta \rangle = B_0 = {}& \theta_f - \chi\sin\theta_\mathrm{LS} \Lambda^B_0, \\
   B_1 = {}& \chi\sin\theta_\mathrm{LS}\exp\left( \Lambda^B_1 \right), \\
   B_2 = {}& -\chi\sin\theta_\mathrm{LS}\exp\left( \Lambda^B_2 \right), \\
   B_3 = {}& \chi\sin\theta_\mathrm{LS}\left(\exp\left( \Lambda^B_3 \right) - 200\right), \\
   B_4 = {}& \left(\Lambda^B_4 \right)^2, \\
   B_5 = {}& -\left(\Lambda^B_5 \right)^2,
\end{align}
where the $\Lambda^X_i$ are a 3-dimension polynomial expansion of the form given by
\begin{align}\label{eqn: global fit expression}
   \Lambda^X_{i} = {}& \sum_{p=0}^3 \sum_{q=0}^3 \sum_{r=0}^4
   \lambda^i_{pqr} \eta^p \chi^q \cos^r\theta_\textrm{LS},
\end{align}
(see Eq. (50) in \PaperOne for reference). Further information on the values of the coefficients $\lambda^i_{pqr}$ can 
be found in Appendix~\ref{sec:waveform-flags}.

\subsubsection{Checks and fallback behavior for $\alpha$ and $\beta$}

As has been previously noted~\cite{Pratten:2020ceb}, in certain parts of the parameter space, the MSA expressions used for $\alpha$ and $\beta$ during inspiral fail.
In the case that these angles fail to generate, we use the next-to-next-to-leading-order (NNLO) angles~\cite{Bohe:2012mr} instead (for explicit expressions see e.g.~\cite{Pratten:2020ceb}), as has been done in previous models which employ the MSA angles~\cite{Pratten:2020ceb}.
The NNLO expressions for the inspiral angles are then treated exactly as described in the preceding subsection-- i.e. rescaling, two-spin mapping and connecting to the merger-ringdown expressions.

Considering $\alpha$, as shown in e.g.~the left-most panel of Fig.~\ref{fig: hm model versus nr}, we see physically incorrect behavior for $A_1 < 0$ ($\alpha$ would decrease as a function of frequency) or $A_2 > 0$ (the dip in $\alpha$ would have the wrong sign).
As it is only a small region of parameter space in which this might happen, we enforce the conditions that $A_1 > 0$ by taking the absolute value of the coefficients with the appropriate sign. For $A_2$ we replace any positive values with zero.
Pathological behavior occurs for $A_3 < 0$.
We limit $A_3$ to have a value above $10^{-5}$ in order to prevent the dip from becoming unphysically narrow.
Finally, in order to ensure that the dip in $\alpha$ does not become unphysically deep, we enforce the following condition; $A_2 < -\pi^2\sqrt{A_3}$.

For $\beta$, we impose the following checks.
First, we use an $\arctan$ window around $\beta$ to ensure it is bounded between $0$ and $\pi$ (see Eq.~(62) and surrounding discussion in \PaperOne for further details).
This ensures that it maintains a physically meaningful value.
To ensure that $\beta$ does not become pathological, if $B_4 \le 175$ then we use $B_4 = 175$.
We determine the correct root of $\beta$ following the prescription set out in Sec.~VI~D of \PaperOne\!.
With the refitting of the coefficients, we also give an updated condition for Eq.~(69) in \PaperOne\!. This condition now reads:
\begin{align}
   \frac{b}{3a} > \frac{B_5}{2} - \frac{2141}{90988}.
\end{align}

The ansatz for $\beta$ can take 3 possible morphologies demonstrated in Fig.~(10) in \PaperOne\!. In the left-hand and center panels, the inflection point occurs 
at a higher frequency than the maximum ($f_\mathrm{inf} > f_\mathrm{max}$). In the center panel, the minimum occurs at a higher 
frequency than the maximum ($f_\mathrm{min} > f_\mathrm{max}$).
In either case, if the maximum occurs above a cutoff frequency $f_\mathrm{low}$, then the connection frequency is given by Eq. (57) 
in \PaperOne\!\!.
\footnote{Eq.~(56) in \PaperOne\!, which defines a quantity $\text{d}\beta_\text{c}$, which appears in Eq.~(57) is incorrect by a factor of 10, i.e. it should be,
\begin{align}
   \text{d}\beta_\text{c} = 25\times10^{-4} \times \text{d}\beta_\text{inf}^2. \nonumber
\end{align}}
Otherwise, the connection frequency is just $f_\mathrm{low}$, which is defined by Eq. (58) in \PaperOne\!.
In the third case (the right hand panel) the connection frequency is defined to be
        \begin{align}
           f_\mathrm{c} =
           \begin{cases}
           f_\mathrm{min} - 0.03 &{} f_\mathrm{min} \ge 0.06  \\
           \frac{3}{5} f_\mathrm{min} &{} f_\mathrm{min} < 0.06
           \end{cases}.
        \end{align}

In the cases shown in the center and right-hand panels of Fig.~(10) in \PaperOne\!, we set the ringdown value of $\beta$ to be constant, given by the value of the merger-ringdown ansatz at that point. This occurs when $f_\mathrm{min} > f_\mathrm{inf}$. For cases like the one on the left-hand panel, we allow $\beta$ to taper to its asymptotic value. To do this we set the connection frequency to $100$.

We only evaluate rescaled inspiral $\beta$ without attaching merger-ringdown contribution in cases where: (i) The lower (inspiral-merger) connection frequency is negative. This is done by setting both connection frequencies to $100$.
(ii) The lower connection frequency is less than 0.0009.
(iii) The value of the merger-ringdown model of $\beta$ is negative at the lower connection frequency.
(iv) We find ourselves choosing the wrong root of the expression for the merger-ringdown value of $\beta$. This is determined by identifying cases where the value of the merger-ringdown model at the lower connection frequency is greater than $5\left(\langle\beta\rangle + 0.5\right)$.

\subsubsection{Outside the calibration region}

Outside the calibration region, we enforce a smooth turnoff of the precession angles so that the model transitions to the inspiral expression for $\alpha$ and the PN-rescaled expression for $\beta$.
This is done using a windowing function of the form of \eqn{pnr-window}, which was
\begin{align}
   w(x) = {}& \frac{1}{2}\left[\cos\left(\frac{x-x_b}{x_w}\right) + 1 \right],
\end{align}
where $x$ is the variable to which we apply the window, $x_b$ is value of the variable at the boundary of the window and $x_w$ is the range of the values of the variable over which the window is applied.
This is the same functional form as is used to turn off the coprecessing tuning outside the calibration region.

For the angles, we smoothly transition to the PN form of the angles once the mass ratio $q$ and single-spin mapped dimensionless spin magnitude $\chi$ are beyond the calibration region.
The final window applied to the model is therefore a product of these two windows (i.e. $w(q)w(\chi)$).
The parameters used in the two windowing functions are
\begin{align}
   q_b = 8.5,  {}& \qquad
   q_w = 3.5, \\
   \chi_b = 0.85, {}& \qquad
   \chi_w = 0.35. \\
\end{align}

The final expression for the angles in this transition window are
\begin{align}
   \alpha = {}& w(q)\,w(\chi)\,\alpha_\mathrm{MR} + \left[1-w(q)w(\chi)\right]\alpha_\mathrm{MSA}, \\
   \beta = {}& w(q)\,w(\chi)\,\beta_\mathrm{MR} + \left[1-w(q)w(\chi)\right]\beta_\mathrm{PN},
\end{align}
where $\alpha_\mathrm{MR}$ and $\beta_\mathrm{MR}$ are the phenomenological expressions for the merger-ringdown angles, $\alpha_\mathrm{MSA}$ is the inspiral expression for $\alpha$ and $\beta_\mathrm{PN}$ is the PN-rescaled expression for $\beta$.

\subsection{Final $\beta$}
\label{sec:final-beta-fit}

In addition to the fit of the angle $\beta$ in the merger-ringdown regime, which implicitly gives the value to which $\beta$ drops after merger, we find it useful to produce an independent fit of this ringdown value.
The value from this independent fit is used in calculating the effective ringdown frequency via Eq.~\eqref{eqn: HOM expression}.
This value is given by a fit
\begin{align}\label{eqn: final beta fit}
    \beta_\mathrm{f} = {}& \theta_f - \chi\sin\theta_\mathrm{LS} \Lambda^B_\mathrm{f},
\end{align}
where $\Lambda^B_\mathrm{f}$ is given by Eq.~\eqref{eqn: global fit expression}. 
From this it can be seen that as with $\langle\beta\rangle$, we use $\theta_\mathrm{f}$ (shown in Fig.~\ref{fig: thetaf behaviour}) to inform the overall shape of $\beta_\mathrm{f}$ and the anti-aligned-spin limit in particular.
Details on the 
fit coefficient values is found in Appendix~\ref{sec:waveform-flags}.

\begin{figure*}[htbp]
   \centering
   \includegraphics[width=1.0\textwidth]{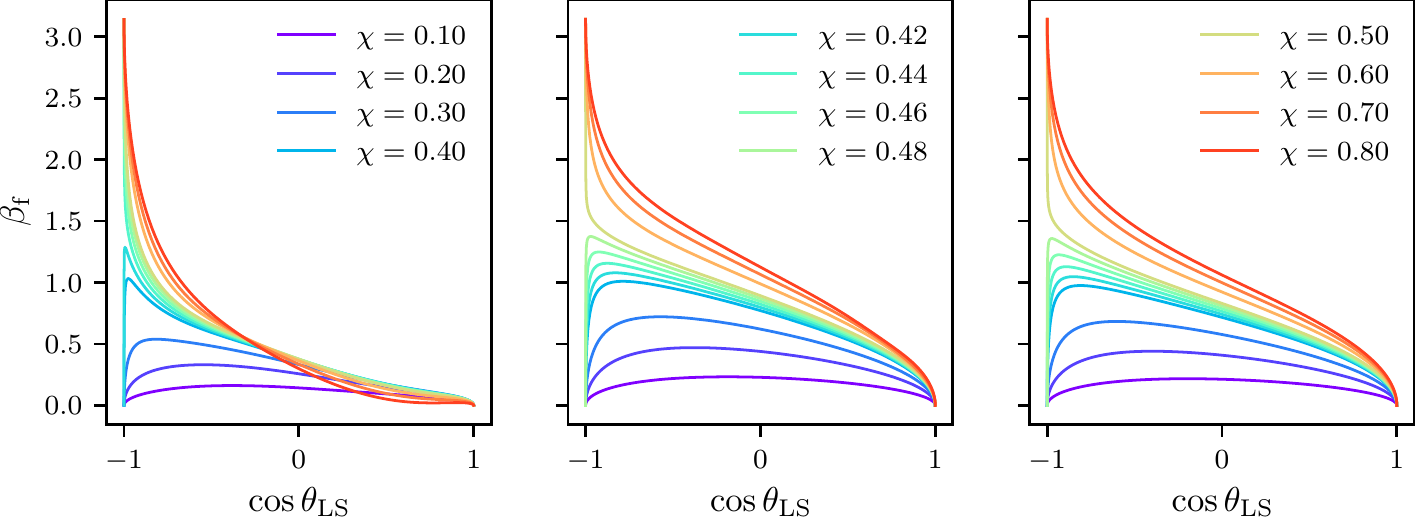}
   \caption{Comparison of $\beta_\mathrm{f}$ values between different models. Left: \modelname{}, center: \tphm{}, right: \vFivePHM{}. Systems under consideration all have $q=8$.}
   \label{fig: final beta}
\end{figure*}

A comparison of the final value of $\beta$ as given by Eq.~(\ref{eqn: final beta fit}) and that employed by other models is shown in Fig.~\ref{fig: final beta}.
We compare here against \vFivePHM{}~\cite{Ramos-Buades:2023ehm} (right hand panel) and \tphm{}~\cite{Estelles:2021gvs} (middle panel) which both set the ringdown value of $\beta$ to a constant equal to the value of the inspiral angle prescription at a specified attachment time.
We do not consider \pxphm{} (MSA), where the inspiral angles are employed through merger and ringdown and do not tend to a constant value or \spintaylor{} (Spin Taylor), which tends smoothly to either 0 or $\pi$ as determined by the \pn{} treatment of the inspiral angles.

It can clearly be seen from this figure that both \vFivePHM{} and \tphm{} tend to overestimate the ringdown value of $\beta$ since they do not capture the rapid drop in the value at merger.
This difference in the values may also be partially accounted for by differences between the time-/frequency-domain values of $\beta$.
However, all three models show roughly the same behavior in the antialigned-spin limit in that they tend either to 0 or $\pi$ depending on the binary configuration.
The exact point at which this transition occurs differs slightly between the models depending on the approximations employed (\modelname{} relies on an approximation of the precessing final spin and remnant quantity fits to \nr{} data whereas \vFivePHM{} and \tphm{} use \pn{} information from the inspiral).
These differences may cause differences when using the various models for data analysis, e.g., parameter estimation for high-mass signals.

\subsection{Additional Changes}
\label{sec:additional-angle-changes}

\subsubsection{Enforcement of the Kerr limit}

The \pxphm{} model includes a check to make sure that the individual spin components do not exceed the physical
Kerr limit. In our mapping of generic spins to an equivalent single-spin configuration, there are regions of parameter space
where the effective single-spin magnitude exceeds one. This is fine: the parameter no longer represents the physical spin
of a single black hole, but an effective spin parameter in the model. As such, the Kerr-limit test
in the code to populate the \texttt{IMRPhenomXPrecessionStruct} is bypassed in the special case of
producing equivalent single-spin precession angles for these cases.

\subsubsection{Re-mapping of \(\theta_\text{JL}\)}
In \texttt{LAL}, the source frame is the frame instantaneously tracking the direction of
the orbital angular momentum at the specified reference frequency~\cite{Schmidt:2017btt}. This frame is related to the
\(\mathbf{J}\)-frame in which the inertial waveform multipole moments are defined by the two angles
\((\theta_\text{JL},\varphi_\text{JL})\), where
\begin{align}
    \label{eq:theta-JN-dynamics}\theta_\text{JL} &= \arccos\left(\mathbf{\hat{J}}\cdot \mathbf{\hat{L}}\right),\\
    \varphi_\text{JL} &= \arctan \left(\frac{S_y}{S_x}\right),
\end{align}
with components defined in the \texttt{LAL} source frame. The expression for \(\varphi_\text{JL}\)
follows from the fact that \(\mathbf{\hat{L}}=\mathbf{\hat{z}}\) in the \texttt{LAL} source frame.

When modeling the \(\mathbf{J}\)-frame signal using our angle model, we are no longer mapping
from a source frame determined by the dynamics but rather by the maximal emission direction,
\(\mathbf{\hat{V}}=\mathbf{\hat{z}}\), in which case the angle \(\theta_\text{JL}\) as computed in
Eq.~\eqref{eq:theta-JN-dynamics} is not necessarily correct. Instead, the angle
\(\theta_\text{JV}=\beta\) should be used to map from the ``maximal emission'' source frame in which
the spins are defined to the \(\mathbf{J}\)-frame.

\subsection{Mapping of precession angles to subdominant multipoles}
\label{sec:hm-angles}

As discussed in Sec.~\ref{sec:frame-choices}, complications arise when attempting
to extend the same modeling assumptions used to produce the quadrupolar precession angles
to higher signal multipoles. The notion that we can simply use the $\ell=2$ Euler angles for the higher
multipoles (as one can in the time domain~\cite{Varma:2019csw}) dissolves, 
and we choose to instead produce a set of angles for each \((\ell,m)\)-multipole included in the model.

\begin{figure}[htbp]
    \centering
    \includegraphics[width=0.45\textwidth]{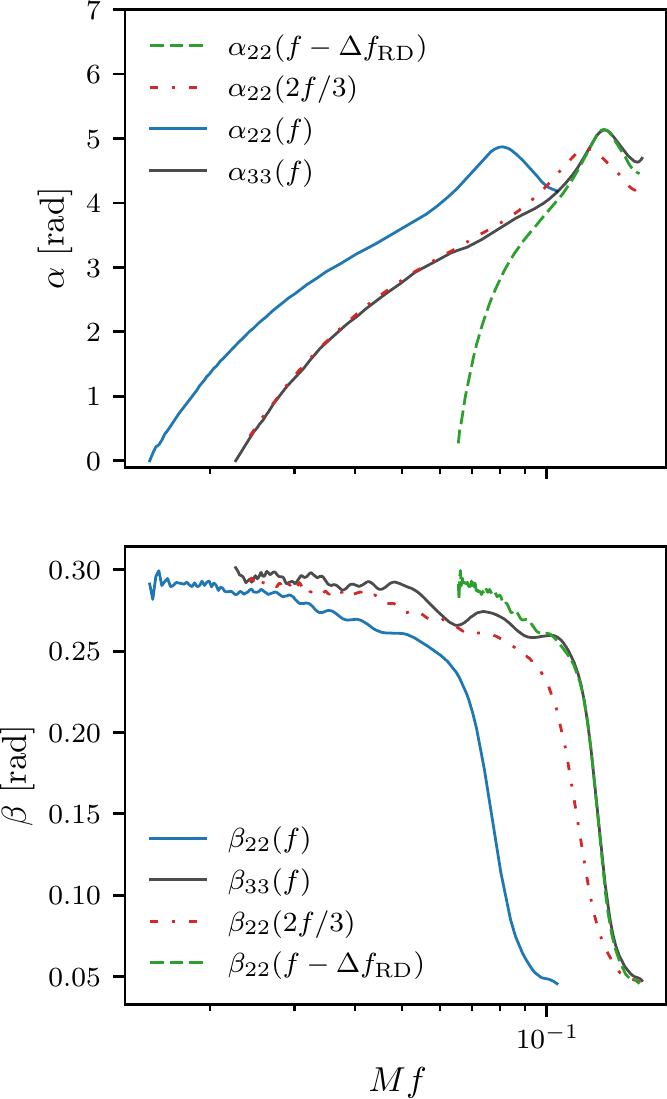}
    \caption{The \(\alpha\) and \(\beta\) precession angles computed from the \((q, \chi,\theta_\text{LS})=(8,0.6,30^\circ)\) simulation shown in the top and bottom panels, respectively. The angles \(\alpha_{\ell \ell}\) and \(\beta_{\ell \ell}\) are computed from the FD strain with only \((\ell,\ell)\)-multipole moments in the coprecessing frame, for \(\ell\in\{2,3\}\). The red dash-dotted curves show the quadrupolar \(\ell=2\) precession angles remaped using the approximate inspiral frequency scaling, whereas the green dashed curves show the quadrupolar precession angles shifted in frequency by the difference of the ringdown frequencies \(\Delta f_\text{RD}=f_{\text{RD},33}-f_{\text{RD},22}\).}
    \label{fig: nr angle rescaling}
 \end{figure}

Furthermore, as seen in Fig.~\ref{fig: nr angle rescaling},
application of the inspiral frequency
rescaling is not sufficient to capture the full
frequency evolution of the higher multipole precession angles. For this
example we compute the FD precession angles
\((\alpha_{22},\beta_{22})\) and \((\alpha_{33}, \beta_{33})\) from the the FD strain
containing only \((\ell,|m|)=(2,2)\) and \((\ell,|m|)=(3,3)\) in the coprecessing frame, respectively.
At low frequencies, the simple frequency rescaling with azimuthal index works well,
demonstrated by the good agreement of the \(\alpha_{22}(2f/3)\)
and \(\beta_{22}(2f/3)\) curves compared directly to \(\alpha_{33}\) and \(\beta_{33}\)
in Fig.~\ref{fig: nr angle rescaling}. This simple map does not work at high
frequencies, where instead the high-frequency behavior of the precession angles
appears to be approximately governed by a shift in the input frequency equal
to the difference of the \((2,2)\)- and \((3,3)\)-multipole ringdown frequencies
in the coprecessing frame. The transition from depending on \(m\) in the inspiral
to depending on \(\ell\) (through the ringdown frequency scaling) in the merger and
ringdown should not come as a surprise, as it follows similarly to
the approximate scalings
seen in the higher signal multipoles~\cite{London:2017bcn}.

We now discuss the specific details in mapping the tuned precession angles
detailed above to the higher multipoles.

\subsubsection{\((\ell,m)\)-angle frequency map}
\label{sec:hm-frequency-map}

To map the tuned, dominant \((2,2)\)-multipole precession angles to rotate the
higher multipoles, we follow a similar approach as detailed in Ref.~\cite{London:2017bcn}.
At low frequencies, the frequency is rescaled with respect to $m$ and the
inspiral contributions to the angle functions are evaluated at the
velocity 
\begin{equation}
   \label{eq: velocity def}
   v = (2\pi M f/m)^{1/3},
\end{equation}
much as is done in previous precessing higher-multipole \phenom{}
models~\cite{Khan:2019kot,Pratten:2020ceb}.

At the higher frequencies of the merger and ringdown, the simple
inspiral rescaling does not hold and we instead
shift the frequency so that each multipole's ringdown
frequency is shifted to the $(2,2)$ ringdown frequency
\begin{equation}
    f \rightarrow f - (f_{\text{RD},\ell m} - f_{\text{RD},22} ).
\end{equation}
The two regions are then connected by a linear mapping between them.

A mapping of the tuned precession angles to rotate the \((\ell,m)\)-multipole
coprecessing signal is created by evaluating the \pnr{} angles using the frequency
map described above, \textit{e.g.},
\begin{equation}
    \label{eq:hm-angle-alpha-map}
    \alpha_{\ell m} (f) \equiv \alpha(f_{22}(f)),
\end{equation}
with
\begin{equation}
    \label{eq:hm-frequency-map}
f_{22}(f) = \left\{
    \begin{array}{ll}
        2f/m & \quad f\leq f_\text{PN} \\
        \xi_1\left(f-f_\text{PN}\right)+\xi_0 & \quad f_\text{PN} < f \leq f_\text{MR} \\
        f - (f_{\text{RD},\ell m} - f_{\text{RD},22}) & \quad f > f_\text{MR},
    \end{array}
\right.
\end{equation}
and linear coefficients,
\begin{align}
    \xi_1&=\frac{f_\text{MR} - (f_{\text{RD},\ell m}-f_{\text{RD,22}}) - 2f_\text{PN}/m}{f_\text{MR}-f_\text{PN}},\\
    \xi_0&=2 f_\text{PN} / m.
\end{align}

\begin{figure*}[htbp]
    \centering
    \includegraphics[width=0.98\textwidth]{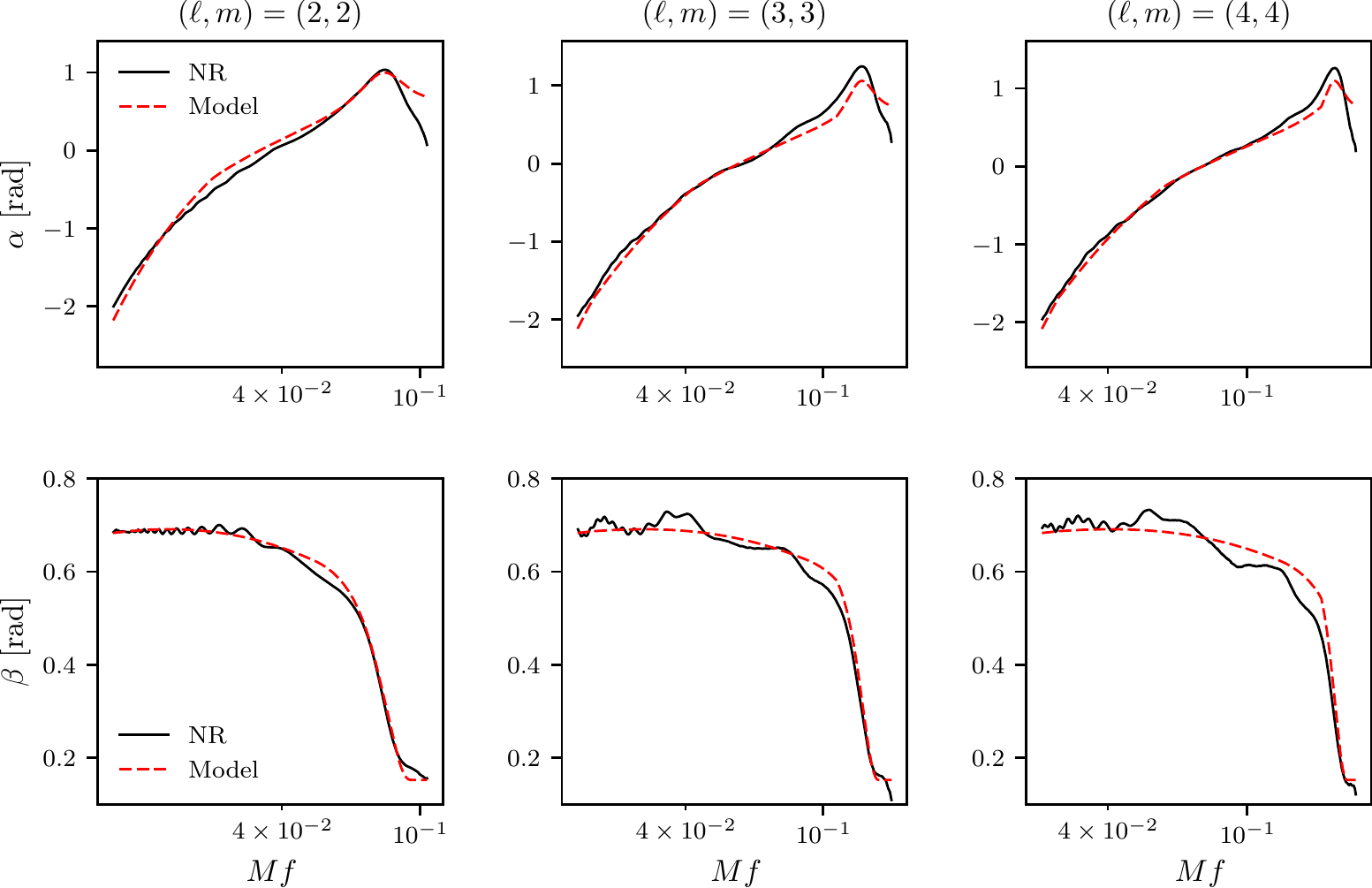}
    \caption{Comparison of the \(\alpha_{\ell \ell}\) and \(\beta_{\ell \ell}\) precession angles computed from the \((q, \chi,\theta_\text{LS})=(4,0.8,90^\circ)\) simulation for \(\ell\in\{2,3,4\}\). }
    \label{fig: hm model versus nr}
 \end{figure*}

We specify the lower and upper connection frequencies, \(f_\text{PN}\) and \(f_\text{MR}\) respectively,
based on connection frequencies defined for both \pnr{} \(\alpha\) and \(\beta\),
\begin{align}
    \label{eq:hm-fmap-intercept}f_\text{PN} &= c_\text{PN}\, m \,f_1 / 2,\\
    \label{eq:hm-fmap-slope}f_\text{MR} &= c_\text{MR} \left(f_\text{c} + f_{\text{RD},\ell m}-f_{\text{RD,22}}\right),
\end{align}
where \(f_1=2A_4/7\) and \(f_\text{c}\) is defined in Eq.~(57) of \PaperOne\!. The
coefficients \(c_\text{PN}\) and \(c_\text{MR}\) are found by minimizing the joint RMS
error between the NR angles and frequency-mapped PNR angles, added in quadrature, for
both $\alpha$ and $\beta$ separately. The values of \(c_\text{PN}=0.65\) and \(c_\text{MR}=1.1\)
were found to be a good initial global fit, and any further tuning with
dependence on intrinsic parameters is left for future work.
An example comparison of the angles \(\alpha_{\ell m}\) and
\(\beta_{\ell m}\) against NR precession angles is shown
in Fig.~\ref{fig: hm model versus nr}.

 We note one subtle issue with the angle frequency map. If we are to re-map the precession angles by analogy with 
 each multipole's frequency evolution, as described above, then for consistency the anti-symmetric (2,2) contribution would be rotated with 
 the same angles as the (2,1) multipole, since that most closely mimics the frequency evolution of the antisymmetric
 (2,2) contribution~\cite{Ghosh:2023mhc}. We experimented with using both the (2,2) and (2,1) angles, and found no 
 appreciable difference in the accuracy with either choice. In this version of the model we have used the (2,2) angles, 
 but this should be reconsidered in future work. 

\subsubsection{\((\ell,m)\)-angle interpolation}
\label{sec:hm-angle-interpolation}
To compute the angles for the higher multipoles as in Eq.~\eqref{eq:hm-angle-alpha-map}
we use the cubic spline interpolant provided by the GNU Scientific Library (\texttt{gsl})~\cite{galassi2009gnu}. This construction requires
us to first produce the tuned \pnr{} precession angles \(\alpha\), \(\beta\), and \(\gamma\) over
the frequency values used in the frequency map in Eq.~\eqref{eq:hm-frequency-map} and with
appropriate frequency spacing.

Given a waveform generated between the frequency values \(f_\text{min}\) and \(f_\text{max}\),
the frequency map in Eq.~\eqref{eq:hm-frequency-map} potentially requires use of the tuned precession
angles outside of the specified frequency range \(f\in[f_\text{min},f_\text{max}]\). To see this one need only look
at the mapping for low frequencies \(f<f_\text{PN}\), where the lowest value is
\(f_{22}(f_\text{min})=f_\text{min}/2\) for the \((4,4)\)-multipole angles. At high frequencies \(f>f_\text{MR}\),
it is possible for \(f_{22}(f_\text{max}) > f_\text{max}\) when mapping to the \((2,1)\)-multipole,
where generally \(f_{\text{RD},21}<f_{\text{RD},22}\). This extension
to the frequency range must be accounted for to avoid extrapolation errors in the spline evaluation.

To appropriately generate interpolants that will cover the required frequency range, we specify modified
minimum and maximum frequency values, \(\breve{f}_\text{min}\) and \(\breve{f}_\text{max}\) respectively,
between which we generate the angles. For the minimum frequency,
we set \(\breve{f}_\text{min} = 2 f_\text{min} / m\) for the largest value of \(m\) contained in
the list of signal multipoles desired in the waveform. To compute the maximum frequency, we specify
\(\breve{f}_\text{max}=f_\text{max} - (f_{\text{RD},21}-f_{\text{RD,22}})\) if the \((2,1)\)-multipole
is desired; otherwise we keep \(\breve{f}_\text{max}=f_\text{max}\).

Finally, once we are equipped with the appropriate frequency spacing \(\Delta f\) (see below), we pad
the minimum and maximum frequencies by \(2\Delta f\) to avoid potential extrapolation due to truncation errors.
Should \(\breve{f}_\text{min} - 2\Delta f < 0\), then \(\breve{f}_\text{min}\) is close to zero and we
instead take half of the minimum frequency \(\breve{f}_\text{min} /2 \).

Generation of the precession angles used to construct the cubic spline interpolants is done on a
uniform frequency grid; we now describe the methods used to estimate an appropriate frequency spacing
for that uniform frequency grid, following loosely the work done on frequency multibanding of the
gravitational wave phasing and precession angles detailed in
Refs.~\cite{Garcia-Quiros:2020qlt,Pratten:2020ceb}.

We initially consider single-spin cases, or cases for which two-spin interactions are negligible, 
and work in units where the total mass \(M=1\).
In these cases, as discussed in Sec.~VD of \cite{Pratten:2020ceb}, the most stringent requirement on \(\Delta f\)
arises from the behavior of \(\alpha\) in the inspiral, which at leading \pn{} order scales with
frequency as,
\begin{equation}
    \label{eq:pn-alpha-scaling}
    \alpha\sim\left(-\frac{5\delta}{64 m_1}-\frac{35}{192}\right)(\pi f)^{-1}.
\end{equation}
The work in~\cite{Garcia-Quiros:2020qlt} considers linear interpolation of the gravitational wave phase
and amplitude, and relates in their Eq.~(2.5) the frequency spacing \(\Delta f\) required to produce an interpolant
with a given error \(R\) to the second derivative of the function being interpolated. In this work
we are using cubic spline interpolants with natural boundary conditions, and we may therefore assume
that the error scaling in this interpolation is approximated by~\cite{HALL1976105},
\begin{equation}
    R(f_*) \leq \max_{f_0\leq f_* \leq f}\frac{5}{384}\alpha^{(4)}(f_*)\Delta f^4,
\end{equation}
for some \(f_0<f_*<f\).

Solving for \(\Delta f\) and using Eq.~\eqref{eq:pn-alpha-scaling}, we find,
\begin{equation}
    \label{eq:delta-f-from-alpha}
    \Delta f = 4\sqrt{\frac{2}{5}}\left( \frac{3 \pi R \left[1+\sqrt{1-4\eta}\right]}{7+13\sqrt{1-4\eta}} \right)^{\frac14}(\breve{f}_\text{min})^{\frac54},
\end{equation}
where we have used \(m_1=(1+\delta)/2,\) \(\delta = \sqrt{1-4\eta}\), and the fact that
the fourth derivative Eq.~\eqref{eq:pn-alpha-scaling} will
be maximized for a given set of parameters at the lowest evaluated frequency \(\breve{f}_\text{min}\).
For the purposes of mapping the higher-multipole angles, we set the
default value of \(R=0.01\), though this error threshold may be modified
using the \texttt{LALSimulation} waveform flag infrastructure (see Appendix~\ref{sec:waveform-flags}).

While the angle model was tuned to single-spin configurations,
the MSA precession angles used
for frequencies covering the inspiral regime describe generic
two-spin configurations and may
contain oscillations induced by changes in the magintude
of the total spin vector. In these cases the above
frequency spacing specified by Eq.~\eqref{eq:pn-alpha-scaling}
may not suffice and we turn to
a different method to predict the required spacing.
The oscillations in the total spin magnitude
are given by~\cite{Chatziioannou:2017tdw},
\begin{equation}
    S_0^2 = S_+^2(t_\text{rr})+\left[S_-^2(t_\text{rr})-S_+^2(t_\text{rr})\right]\text{sn}[\psi(t_\text{pr},t_\text{rr}),m(t_\text{rr})],
\end{equation}
where \(S_-^2\) and \(S_+^2\) are the squared magnitudes of the minimum and maximum
total spin vector configurations, respectively, \(t_\text{rr}\) is the radiation-reaction timescale,
\(t_\text{pr}\ll t_\text{rr}\) is the precession timescale, \(\psi\) is the phase angle tracking
the oscillation between \(S_-\) and \(S_+\), and \(m=(S_+^2-S_-^2)/(S_+^2-S_3^2)\).

\begin{figure*}[htbp]
    \centering
    \includegraphics[width=0.98\textwidth]{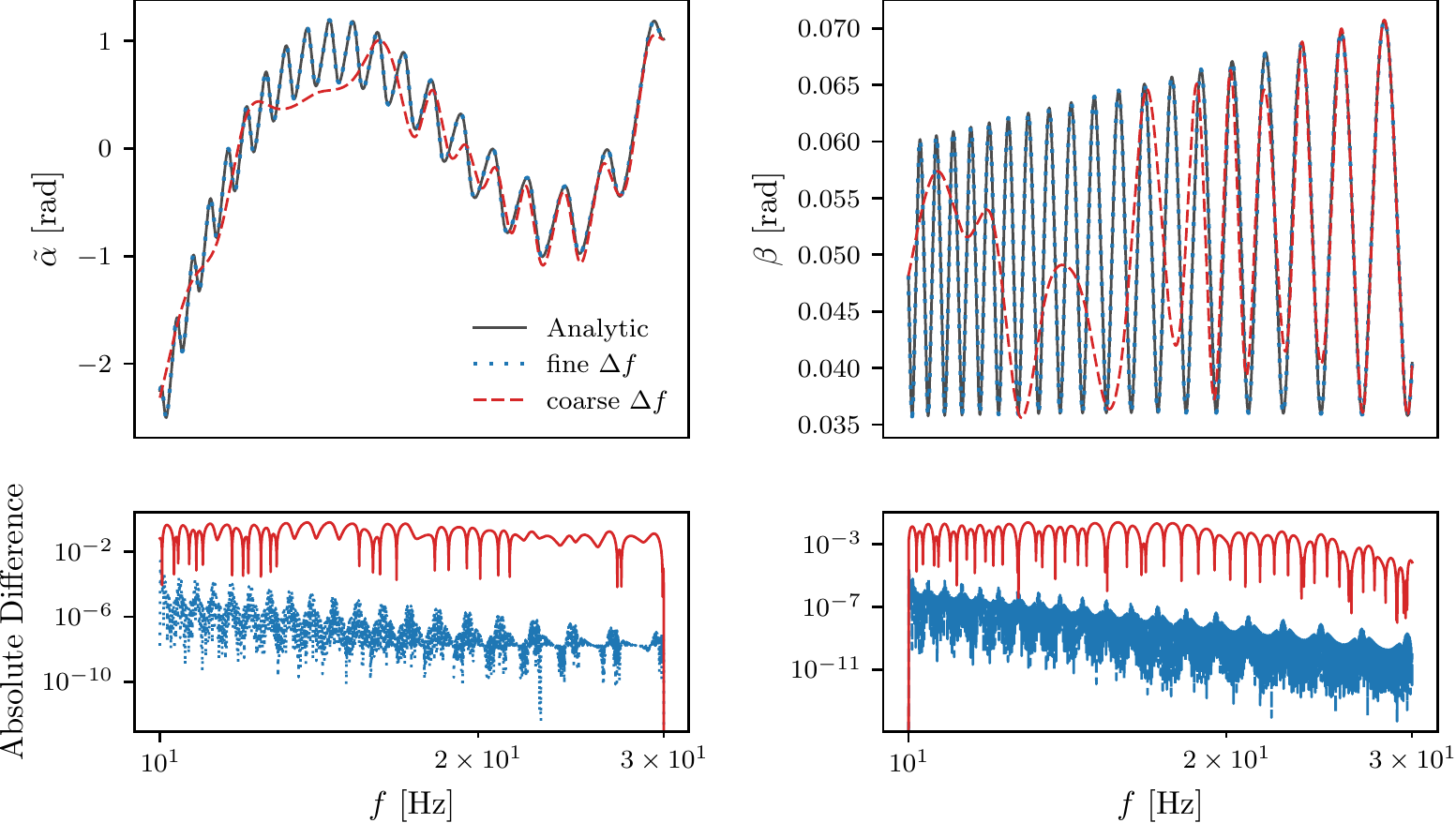}
    \caption{Comparison between the analytic evaluation of \(\alpha\) and
    \(\beta\) and two interpolants generated with different frequency
     resolutions, ``fine \(\Delta f\)'' from Eq.~\eqref{eq:delta-f-from-oscillations} and ``coarse \(\Delta f\)'' from Eq.~\eqref{eq:delta-f-from-alpha}, for the case detailed in
     Table~\ref{tab: two-spin config}. The top panels show plots of the
     precession angles, where we have defined \(\tilde\alpha\) to be the
     difference between \(\alpha\) and a quadratic polynomial fit to the analytic \(\alpha\) to better facilitate visualiztion of the two-spin oscillations. The bottom panels show the absolute difference between the analytic angles and the interpolants when evaluated on the same frequency grid as the analytic angles. Here the red solid curves correspond to the difference between the analytic angles and coarsely-sampled interpolants, and the blue dashed curves show the difference between the analytic angles and finely-sampled interpolants.
     }
    \label{fig: interpolation error}
 \end{figure*}
 
The solution for \(\psi\) is given by Eq.~(51) of~\cite{Chatziioannou:2017tdw},
\begin{equation}
    \label{eq:spin-magnitude-phase}
    \psi = \psi_0 - \frac{3 g_0 \delta m}{4} v^{-3} \left(1 + \psi_1 v + \psi_2 v^2 \right),
\end{equation}
where \(\psi_0\) is a constant of integration, \(v\) is the velocity defined in 
Eq.~\eqref{eq: velocity def}, and the remaining terms are defined
in~\cite{Chatziioannou:2017tdw}. 
We will assume that over a small range
of frequency
values close to \(\breve{f}_\text{min}\), we can approximate
\begin{equation}
    \label{eq: psi approximation}
    \text{sn}[\psi(\breve{f}_\text{min}),m] \approx \sin(\psi' \breve{f}_\text{min}),
\end{equation}
where \(\psi'\) is the first derivative of Eq.~\eqref{eq:spin-magnitude-phase} with respect to \(f\),
\begin{equation}
    |\psi'| = \frac{\pi g_0 \delta m }{4 v^6}\left(3 + 2 \psi_1 v + \psi_2 v^2 \right).
\end{equation}
Then to adequately resolve the oscillations in Eq.~\eqref{eq: psi approximation} we choose to specify a sampling rate that places four
frequency points within one period of oscillation, \textit{i.e.},
\begin{equation}
    \label{eq:delta-f-from-oscillations}
    \Delta f = \frac{1}{4|\psi'|}.
\end{equation}
 
This approximation handles almost all cases of two-spin oscillations. For some configurations where
the minimum and maximum values of $\beta$ satisfy \(0\approx \beta_\text{min} < \beta_\text{max}\)
we see sweeping oscillations in $\beta$ that drop close to zero, at which point
the coordinates of the rotation approach a singular point
and $\alpha$ sees sharp jumps of \(\pi\). To predict these cases we compute the minimum and maximum
values that $\beta$ can take following~\cite{Pratten:2020ceb},
\begin{align}
    \beta_\text{min} &= \arctan \left( \frac{|S_{1\perp} - S_{2\perp}|}{L + S_\parallel}\right),\\
    \beta_\text{max} &= \arctan \left( \frac{S_{1\perp} + S_{2\perp}}{L + S_\parallel}\right),
\end{align}
where \(L\) is the magnitude of the post-Newtonian orbital angular momentum used by \pxphm{} evaluated
at \(\breve{f}_\text{min}\), \(S_\parallel\) is the component of the total spin parallel to \(\mathbf{L}\),
and \(S_{1,2\perp}\) are the components of \(\mathbf{S}_1\) and \(\mathbf{S}_2\) perpendicular to
\(\mathbf{L}\), respectively.
 
When the conditions \(\beta_\text{min} < 0.01\) and \(\beta_\text{min} / \beta_\text{max} < 0.55\) are both
met, then we can assume that $\beta$ is oscillating sufficiently close to zero for jumps in $\alpha$ to
be a potential concern. In this case, we increase the resolution by dividing \(\Delta f\) in
Eq.~\eqref{eq:delta-f-from-oscillations} by a factor of 4,
\begin{equation}
    \label{eq:delta-f-singular}
    \Delta f = \frac{1}{16|\psi'|}.
\end{equation}
Finally, we use the minimum \(\Delta f\) computed from either Eqs.~\eqref{eq:delta-f-from-alpha},
~\eqref{eq:delta-f-from-oscillations},~or~\eqref{eq:delta-f-singular}, and require
that the final choice of \(\Delta f \ge 10^{-2}\) so as not to
saturate available memory when generating the waveform.

\begin{table}
    \begin{tabular}{l | c}
        \hline
        Parameter & Value\\
        \hline
        $M$ [$M_\odot$] & 34.7488472
        \\
        $q$ & 15.1486886
        \\
        $\chi_1$ & $(-0.04585042, -0.03174999,  0.65327636)$
        \\
        $\chi_2$ & $(-0.90498966,  0.31916086, -0.06208559)$
        \\
        \hline
    \end{tabular}

    \caption{Configuration used to generate Fig.~\ref{fig: interpolation error}. The angles were produced with starting and reference frequencies of 10 Hz.}
    \label{tab: two-spin config}
\end{table}

In Fig.~\ref{fig: interpolation error} we show the impact of inadequate
frequency spacing on the interpolation of the oscillating
precession angles. The precession angles \(\alpha\) and \(\beta\)
are presented in the top two panels of the figure for the configuration listed in Table~\ref{tab: two-spin config}, where the analytic
evaluation of the angles with \(\Delta f=1\times 10^{-3}\)~Hz is shown in
black solid curves. The blue dotted and red dashed lines show the
results of inerpolating the analytic \(\alpha\) and \(\beta\) using
the frequency step
sizes given in Eqs.~\eqref{eq:delta-f-from-oscillations} (\(\Delta f=0.027\)~Hz) and
\eqref{eq:delta-f-from-alpha} (\(\Delta f=0.843\)~Hz), respectively. To more clearly show the
oscillatory behavior of
\(\alpha\), we define \(\tilde\alpha\) to be
\(\alpha\) minus a quadratic polynomial fit to the analytic evaluation
of \(\alpha\), thereby removing the dominant \(f^2\) growth of
the angle with frequency.

Unsurprisingly, the oscillations in \(\alpha\) and \(\beta\) require finer
frequency spacing to be accurately resolved compared to the simple
leading-order-in-frequency estimate provided
from Eq.~\eqref{eq:delta-f-from-alpha}. This can be seen quite
clearly from the absolute differences between the interpolants
and the analytic angles displayed in the lower panels of
Fig.~\ref{fig: interpolation error}. As the period of the two-spin
oscillations eventually increases enough for the coarse
frequency resolution
to resolve the oscillations (at roughly the highest
frequencies plotted in Fig.~\ref{fig: interpolation error}),
the error from the coarsely-sampled interpolation begins to drop.

\section{Model Performance and PE Results}
\label{sec:model-performance}

We now look to assess the performance of \modelname{} by means of both mismatches and \pe. We adopt
the definitions of the mismatch \(\mism\) and precessing mismatch \(\pmism\) discussed in Sec.~XI\,A of 
\PaperOne\!\!. In computing mismatches we
utilize the advanced LIGO power spectral density at design sensitivity~\cite{aligopsd}.

In Sec.~\ref{sec:coprec-matches}
we first analyze the perfomance of the underlying coprecessing \xcp{} model and also assess the impact of adding
the antisymmetric contribution. We then extend our analysis to look at precessing mismatches with symmetrized
\((\ell,|m|)=(2,2)\) coprecessing data in Sec.~\ref{sec:sym-l2-matches} and full \(\ell=2\)
coprecessing data in Sec.~\ref{sec:l2-matches}. Precessing mismatches using all available coprecessing frame
multipoles and the mapped HM-angles are presented in Sec.~\ref{sec:full-matches}.
Finally we give results from \pe{} recoveries performed with \modelname{} alongside other contemporary models in
Sec.~\ref{sec:pe}.

\subsection{Coprecessing frame mismatches}
\label{sec:coprec-matches}

\begin{figure*}[htbp]
   \centering
   \includegraphics[width=0.98\textwidth]{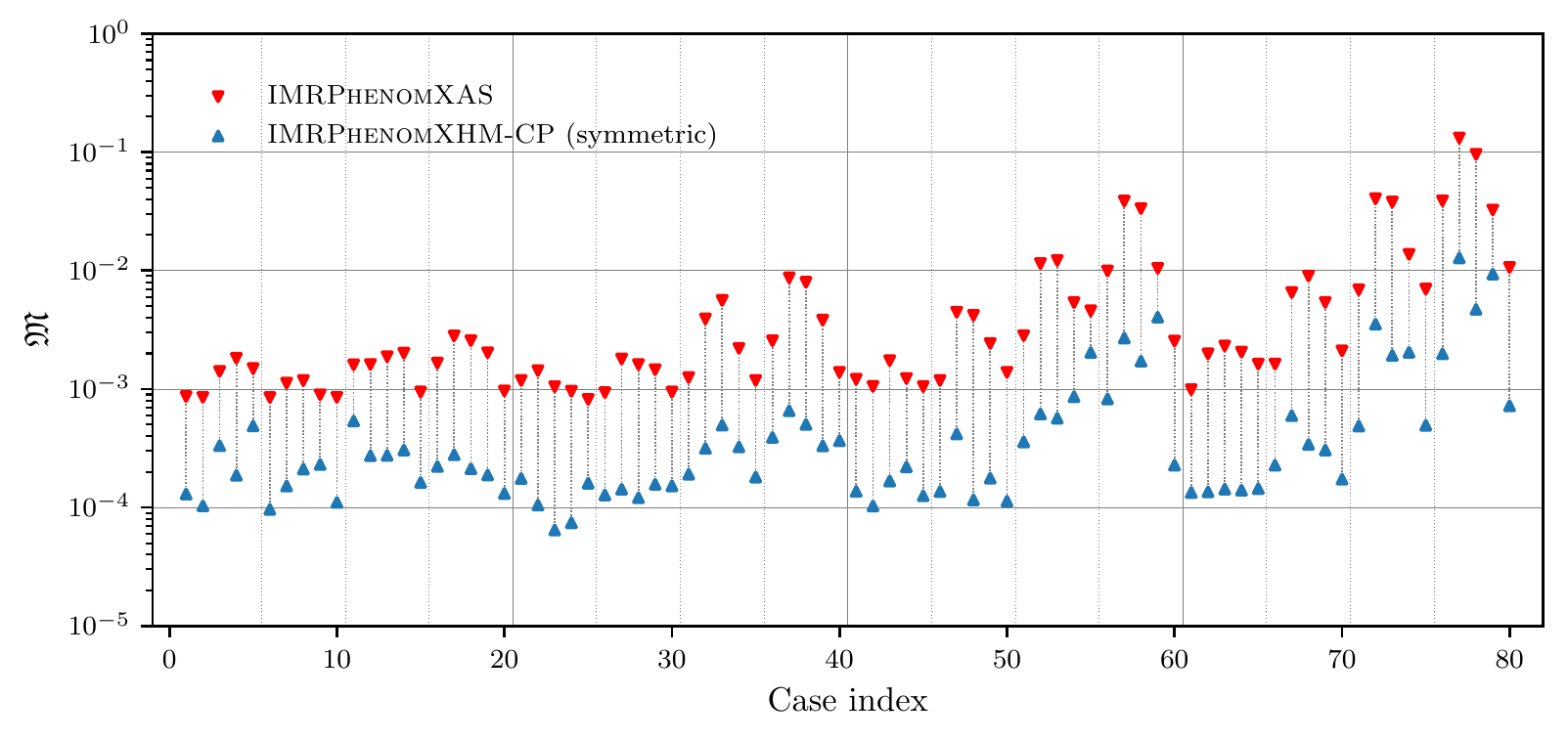}
   \caption{Mismatches \(\mism\) computed between the symmetrized \((\ell,m)=(2,2)\) coprecessing \nr{} data tabulated in Table~I of Ref.~\cite{Hamilton:2021pkf} and the symmetric \((\ell,m)=(2,2)\) coprecessing strain generated with the models \xcp{}, shown with upward pointing arrows, and \pxas{} shown with downward pointing arrows.}
   \label{fig:22-coprec-sym-matches}
\end{figure*}

\begin{figure*}[htbp]
   \centering
   \includegraphics[width=0.98\textwidth]{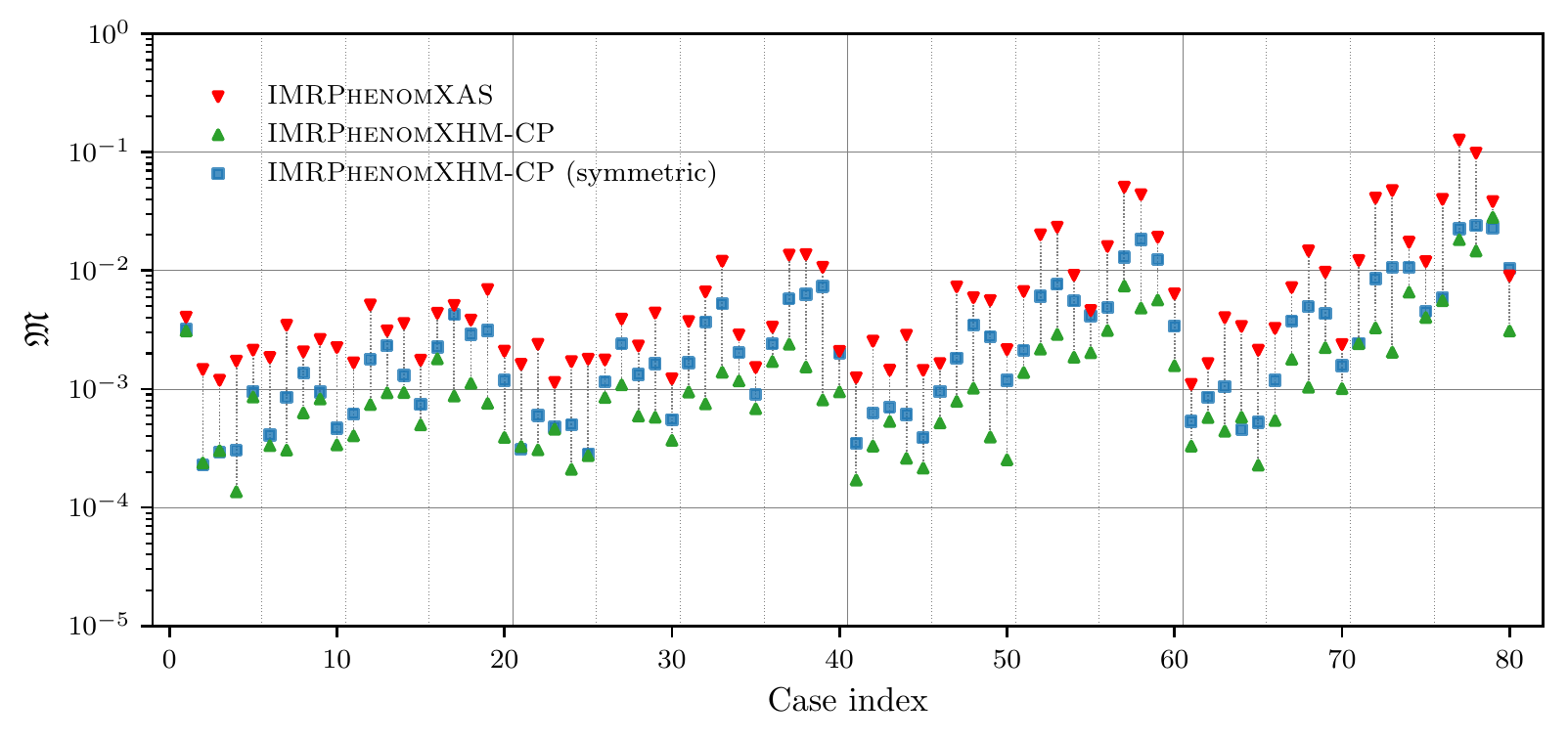}
   \caption{Mismatches \(\mism\) computed between the full \((\ell,m)=(2,2)\) coprecessing \nr{} data with antisymmetric contributions tabulated in Table~I of Ref.~\cite{Hamilton:2021pkf} and the \((\ell,m)=(2,2)\) coprecessing strain generated with the models \xcp{}, shown with upward pointing arrows, and \pxas{} shown with downward pointing arrows. Here the \nr{} data is not symmetrized and the antisymmetric contribution to \xcp{} is enabled. We also include the match between the full coprecessing \nr{} and \xcp{} with the antisymmetric contributions to the model disabled, shown as blue squares. The improvement in performance achieved by the addition of the antisymmetric part to the co-precessing waveform is evident.}
   \label{fig:22-coprec-asym-matches}
\end{figure*}

We start by computing mismatches between the tuned \((\ell,|m|)=(2,2)\) 
symmetric coprecessing model contribution to
\xcp{} and the 80~\nr{} waveforms detailed in Ref.~\cite{Hamilton:2023qkv} and compare the results to the modified
aligned-spin verson of \pxas{} used in \pxphm{}. All mismatches in this subsection use a total
mass of \(100\,M_\odot\), starting frequency \(f_\text{min}=\text{min}\,\left\{20\,\text{Hz},
f_\text{ref}+5\,\text{Hz}\right\}\) and \(f_\text{max}=1024\,\text{Hz}\).

As an initial result shown in Fig.~\ref{fig:22-coprec-sym-matches}, we verify that the
performance improvement seen in \PaperOne by tuning the dominant
quadrupolar contribution to the underlying coprecessing model in \pdcp{} is retained in \xcp{}. Across the
calibration waveforms we see improvement in the mismatch performance over the coprecessing model used
in \pxphm{} for the dominant \((\ell,|m|)=(2,2)\) contribution. This relative improvement is maintained
when comparing to the full \((\ell,m)=(2,2)\) \nr{} coprecessing data without symmetrization,
shown in Fig.~\ref{fig:22-coprec-asym-matches}. While the mismatches for both
\xcp{} and \pxphm{} degrade slightly in overall performance, a majority of the \xcp{} mismatches remain
below~\(10^{-3}\). For this set of matches, the antisymmetric contributions to \xcp{} are enabled, and 
we optimize the matches over rotations of the in-plane spin direction, thereby optimizing over the antisymmetric phase.  Figure~\ref{fig:22-coprec-asym-matches} nicely demonstrates the improvement in the coprecessing model due to modeling the dominant multipole asymmetry. Notably, for about a third of the cases, mismatches $\sim$$10^{-3}$ were only attainable after adding the model of the antisymmetric waveform to the coprecessing model.

Finally in this section we remark on the performance of the higher multipole contributions to \xcp{}
described in Sec.~\ref{sec:hm-coprec}, where no tuning has been done but the ringdown
frequencies used in each multipole are modified by Eq.~\eqref{eqn: HOM expression}. Mismatches between 
individual coprecessing higher multipoles were computed for all 80~NR waveforms (except the \(q=1\) configurations
for the odd-\(m\) multipoles, where the coprecessing contributions approximately vanish). In general the 
performance improvement is lower compared to the results shown in Fig.~\ref{fig:22-coprec-sym-matches},
which is to be expected as the full tuning to NR has not been done. On average the mismatches improve 
across the parameter space when using the effective ringdown frequency for each multipole, with larger improvements seen at higher mass-ratios and spin magnitudes, and we 
report the mean percentage improvement in the mismatches for each \((\ell, |m|)\)-multipole: \((2,1)\): \(0.09\%\), 
\((3,3)\): \(0.17\%\), \((3,2)\): \(0.05\%\), \((4,4)\): \(0.14\%\).

We expect that explicit tuning of the higher multipoles could achieve similar levels of accuracy to that
achieved for the co-precessing-frame (2,2) multipoles. However, this also requires that the model accurately
capture the relative phasing between the multipoles. We will return to this point when we consider full 
precessing matches in Sec.~\ref{sec:full-matches}.

\subsection{Symmetrized $h^\mathrm{CP}_{22}$ mismatches}
\label{sec:sym-l2-matches}

We next consider the accuracy of both our underlying symmetric dominant-coprecessing-multipole and of the dominant multipole precession angles to the waveform. As described above, we have calibrated both the symmetric coprecessing (2,2) multipole and the merger-ringdown part of the precession angles $\alpha$ and $\beta$ to a set of 80 \nr{} waveforms~\cite{Hamilton:2023qkv}. In order to test the performance of this model and the validity of the approximations made, we calculate the full SNR-weighted precessing match between the $\ell=2$ multipoles of the model in the inertial frame and the 80 calibration waveforms.

To do this, we use the cleaned and symmetrized \nr{} waveforms used in the calibration process, more details of which can be found in \PaperOne\!.
The model data are produced by calling \modelname{} with only the $(2,2)$ multipole activated in the coprecessing frame and the multipole asymmetries turned off.
For comparison, we also consider \vFivePHM{}, which is also called with only the $(2,2)$ multipole activated (and natively does not include multipole asymmetries).

When calculating the match, we consider masses in the range between $100\mathrm{M}_\odot$ and $240\mathrm{M}_\odot$ at intervals of $20\mathrm{M}_\odot$.
We calculate the match at each of the inclination values in the set $\left\{0, \pi/6, \pi/3, \pi/2\right\}$.
The match is performed over the frequency range
$f_\textrm{min} = \mathrm{max}\{20\,\mathrm{Hz}, f^\mathrm{NR}_\textrm{min}+5\mathrm{Hz}\}$ to $f_\mathrm{max} =512$~Hz.
This set up is chosen to allow for direct comparison with results in Sec. XI E in \PaperOne utilizing \pnr{}.

\begin{figure*}[htbp]
   \centering
   \includegraphics[width=0.98\textwidth]{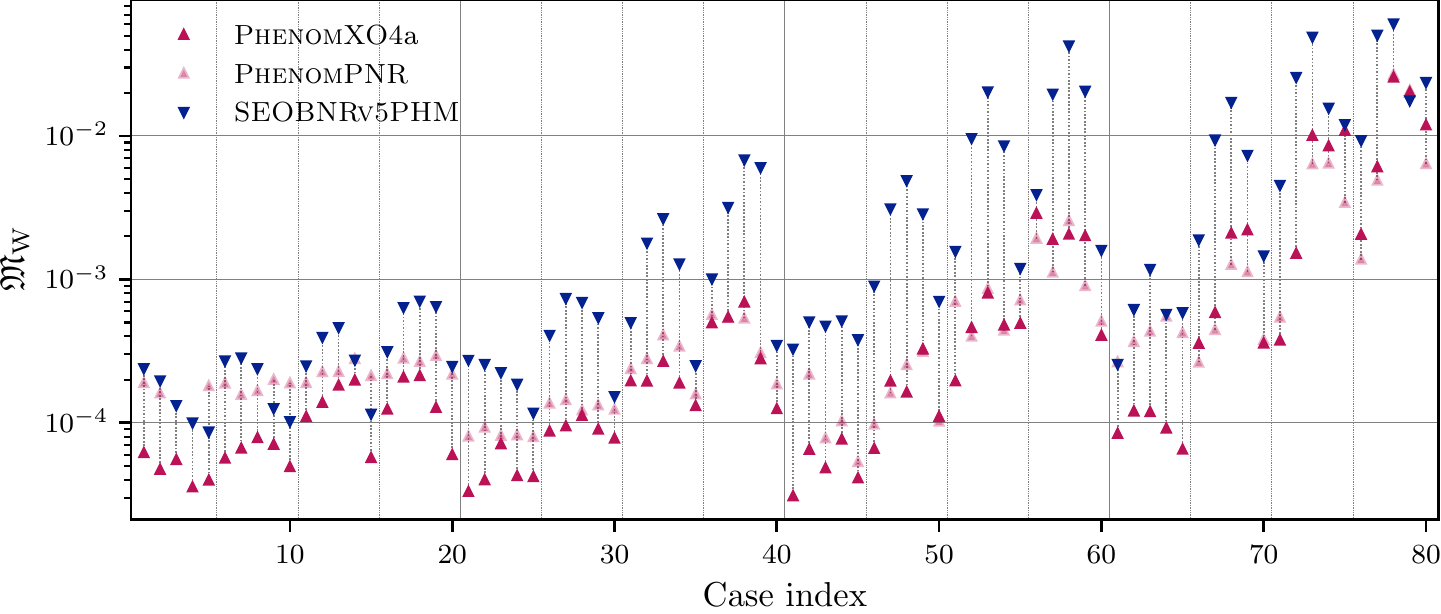}
   \includegraphics[width=0.98\textwidth]{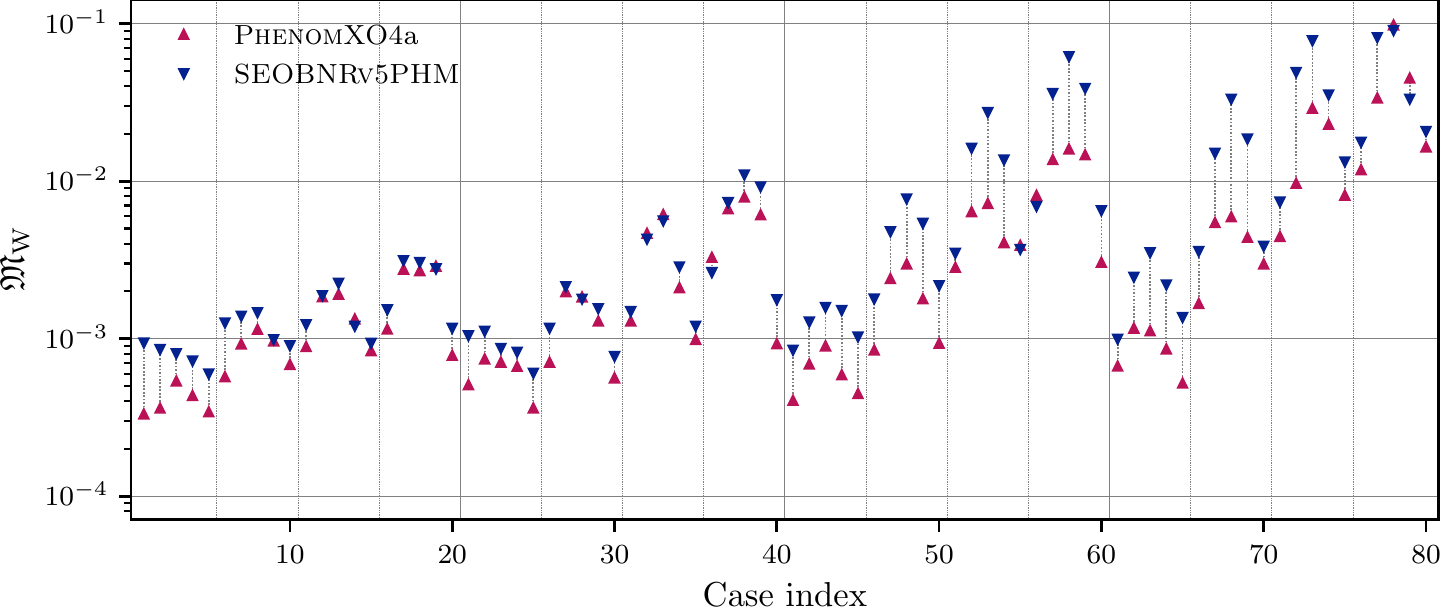}
   \includegraphics[width=0.98\textwidth]{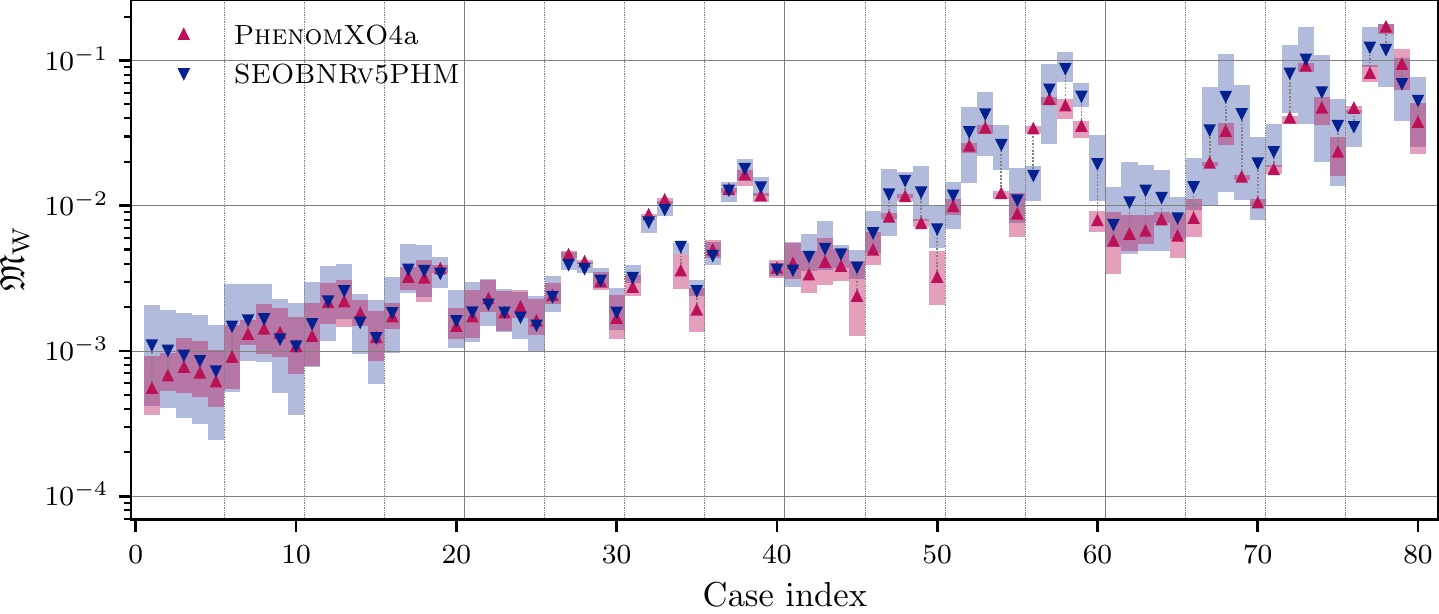}
   \caption{SNR-weighted mismatches in the $\mathbf{J}$-frame between the \nr{} BAM waveforms and the \modelname{} and \vFivePHM{} models containing (a) only the $(2,2)$ multipole in the coprecessing frame -- against symmetrized \nr{} data (top panel) (b) all available coprecessing frame $\ell=2$ multipoles (middle panel) and (c) all available coprecessing frame multipoles up to $\ell=4$. In the top panel, a comparison with \pnr{} is included for reference. The points are averaged over total mass and inclination; the bars show the spread with total mass.}
   \label{fig: BAM match}
\end{figure*}

The results of this comparison can be seen in the top panel of Fig.~\ref{fig: BAM match}.
First, we can see that the improvements to the calibration presented in this paper show an improvement of \modelname{} over \pnr{} on average across the parameter space.
In particular, we see an improvement in the low-spin regime due to the improved treatment of the zero-spin limit in the modeling of the precession angles.
It is important to note that \pnr{} was originally calibrated to a subset of just 40 of the numerical waveforms, whereas \modelname{} has been calibrated to the entire set.
Fitting to just a subset of the waveforms with \pnr{} enabled us to check that there was no overfitting or other issues. 
We then calibrated \modelname{} against the additional 40 waveforms to further improve accuracy.
We note that the performance of \pnr{} for the additional spin magnitudes of $\{0.2, 0.6\}$ is consistent with the accuracy achieved for the calibration set (i.e. $\chi=[0.4,0.8]$). This indicates that the good matches seen here are not purely a consequence of comparing to the calibration set of waveforms.

We can also see the effect of the calibration on these matches when compared to \vFivePHM{}, a recent state-of-the-art precessing model which does not calibrate precession effects to \nr{}.
\modelname{} performs better than \vFivePHM{} across the parameter space, often by an order of magnitude.
A comparison for these cases with the older models \pxphm{} and \vFourPHM{} can be seen in \PaperOne\!.
We see the same improvement there when comparing a model with calibrated precession effects against those without.

\subsection{\(\ell=2\) mismatches}
\label{sec:l2-matches}

We now consider the impact of the inclusion of higher order multipoles and asymmetry on the model accuracy.
We first include all $\ell=2$ coprecessing multipoles used in each model.
\modelname{} and \vFivePHM{} contain the symmetric $(2,\pm2)$ and $(2,\pm1)$ multipoles in the coprecessing frame, and 
\modelname{} also includes the antisymmetric $(2,\pm2)$ multipoles. 
The \nr{} waveforms contain all $\ell=2$ multipoles in the inertial frame without further processing (i.e. no symmetrization).
Once again we calculate the full SNR-weighted precessing match in the inertial frame.
We consider systems with total mass $\{60,90,120,150\}M_\odot$ and inclinations $\{0,\pi/3, \pi/2, 2\pi/3, \pi\}$.
The match is performed over the frequency range $f_\textrm{min} = \mathrm{max}\{20\,\mathrm{Hz}, 1.35 f^\mathrm{NR}_\textrm{min}\}$ to $f_\mathrm{max} =512$~Hz.
This is a slightly different set up to that considered in the previous subsection.
Since we are now considering the effect of asymmetry on the waveform, we now include inclinations greater than $\pi/2$.
As we are examining a greater number of systems than in \PaperOne we sample the total mass parameter space less densely in order to ensure that the analysis is computationally feasible.

First we consider the performance of \modelname{} against the set of 80 calibration BAM waveforms, shown in the middle panel of Fig.~\ref{fig: BAM match}. Here we show the mismatch value averaged over all masses and inclinations. 
Since these results have been produced using a slightly different set of choices for the total masses and inclinations, they cannot be compared directly with the results for the symmetrized $(2,2)$-only matches shown in the top panel. However, we can see that the mismatch degrades by up to an order of magnitude with the addition of the untuned $(2,1)$ multipoles in the coprecessing frame and when considering the multipole asymmetries.
It is unclear which of these additions has the dominant impact on the mismatches.
For systems with spin magnitude below 0.4, we see that the mismatches lie below $10^{-2}$ for all cases up to $q=8$.
For systems with $q\ge4$ and spin magnitudes above 0.4, the mismatch value starts to become notable.
It is therefore important to improve the calibration of the model in this part of the parameter space.
We also consider the performance of \vFivePHM{} against the same set of waveforms.
When considering just the $\ell=2$ multipoles, \modelname{} still outperforms \vFivePHM{} in almost all cases. This improved performance is particularly significant for the higher mass ratio cases included in the dataset.

\begin{figure}[htbp]
   \centering
   \includegraphics[width=0.48\textwidth]{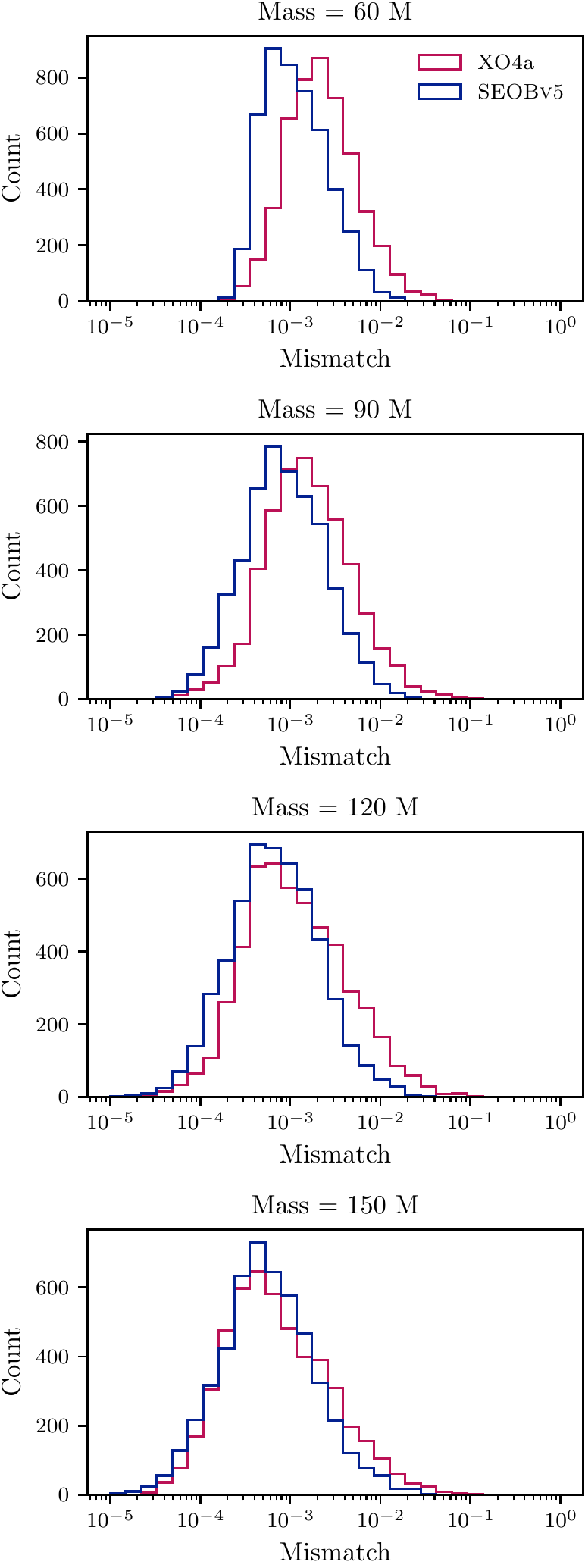}
   \caption{Mismatches between the $\ell=2$ multipoles in the $\mathbf{J}$ frame of  \nrsurSDQF{}, \modelname{} and \vFivePHM{}. Considered 1024 two-spin configurations at four different total masses $\{60,90,120,150\}M_\odot$ and five different inclinations $\{0,\pi/3,\pi/2,2\pi/3,\pi\}$ rad.}
   \label{fig: NRSur ell2 2spin}
\end{figure}

It is worth noting here the uncertainty in the \nr{} waveforms. The mismatch uncertainty in the BAM waveforms is estimated to be 
$\sim$$10^{-3}$~\cite{Hamilton:2023qkv}. We have also calculated mismatches between BAM waveforms and SXS waveforms where equivalent 
configurations exist, and those results are consistent with the same level of disagreement. We therefore caution against interpreting any significance to 
mismatch differences that are smaller than this threshold, which is in general the level of improvement between \modelname{} and \vFivePHM{}
$\ell=2$ mismatch results in the middle panel of Fig.~\ref{fig: BAM match} until we reach the cases where $q = 4$ and $\chi = 0.6$ 
(above case 50). For example, in mismatch calculations where we replace the BAM waveforms by equivalent \nrsurSDQF{} waveforms, 
we find that the relative performance of \modelname{} and \vFivePHM{} often swaps, but changes are not more than roughly $10^{-3}$. 

We also consider single-spin \nrsurSDQF{} configurations that differ from those in our calibration set, i.e., with spin magnitudes and 
misalignment angles in between those used for the BAM simulations. We find similar levels of mismatch to neighbouring configurations,
again demonstrating that over-fitting is not a significant source of error. 

We also consider the performance of these two models against a set of 1024 two-spin configurations generated using \nrsurSDQF{}. 
Due to the region of validity of the surrogate model, these configurations only extend up to mass ratio $q=4$.
These results are shown in Fig.~\ref{fig: NRSur ell2 2spin}.
From this comparison we can see that in general \vFivePHM{} slightly outperforms \modelname{}.
A closer inspection of the results shown in Fig.~\ref{fig: NRSur ell2 2spin} reveals that the performance is roughly comparable between the two models, with a tail extending towards higher mismatches present for \modelname{} accounting for most of the difference.
This effect is less prevalent at higher total mass, where the signal consists of mainly merger and ringdown.
From this we can infer that it is likely the inspiral part of \modelname{} which requires the greatest improvement while the merger-ringdown part has already been accurately tuned.

There are a number of possible reasons for the different pictures of the model accuracy shown in 
Figs.~\ref{fig: BAM match} and~\ref{fig: NRSur ell2 2spin}. 
First and foremost is the presentation of the data --- in Fig.~\ref{fig: BAM match} we average over inclination and total mass, while in Fig.~\ref{fig: NRSur ell2 2spin} we do not, so extremely good/bad cases are more prominent.
Secondly, the mismatch uncertainty between BAM and NRSur of $\mathcal{O}(10^{-3})$ is reflected in the performance of the models against three different ``datasets." Finally, the random sampling of the two-spin parameter space in Fig.~\ref{fig: NRSur ell2 2spin} very rarely samples the 
large-mass-ratio large-primary-spin cases for which \vFivePHM{} shows poor matches in Fig.~\ref{fig: BAM match}.
The picture is however consistent when examining calibration and non-calibration cases, so we do not attribute the apparent change in performance to any overtuning of the model. From this we conclude that \vFivePHM{} slightly outperforms \modelname{} in the bulk of the parameter space, but
is less accurate for more extreme configurations.

\subsection{HM Mismatches}
\label{sec:full-matches}

Finally, we consider the overall performance of the final model. We use the same match set up as described in the preceding section. We consider a data set consisting of the 80 calibration BAM waveforms and 1024 two-spin configurations generated using \nrsurSDQF{}. We compare \modelname{} with the other contemporary time- and frequency-domain waveform models.

The results of the comparison against the 80 calibration BAM waveforms are shown in 
the bottom panel of Fig.~\ref{fig: BAM match}. We can see from a direct comparison 
with the middle panel (which shows the mismatch results for just the $\ell=2$ multipoles) 
that the inclusion of the higher order multipoles further degrades the performance of 
both \modelname{} and \vFivePHM{}, and we now see 
that the performance of the two models is roughly comparable.

\begin{figure*}[htbp]
   \centering
   \includegraphics[width=0.98\textwidth]{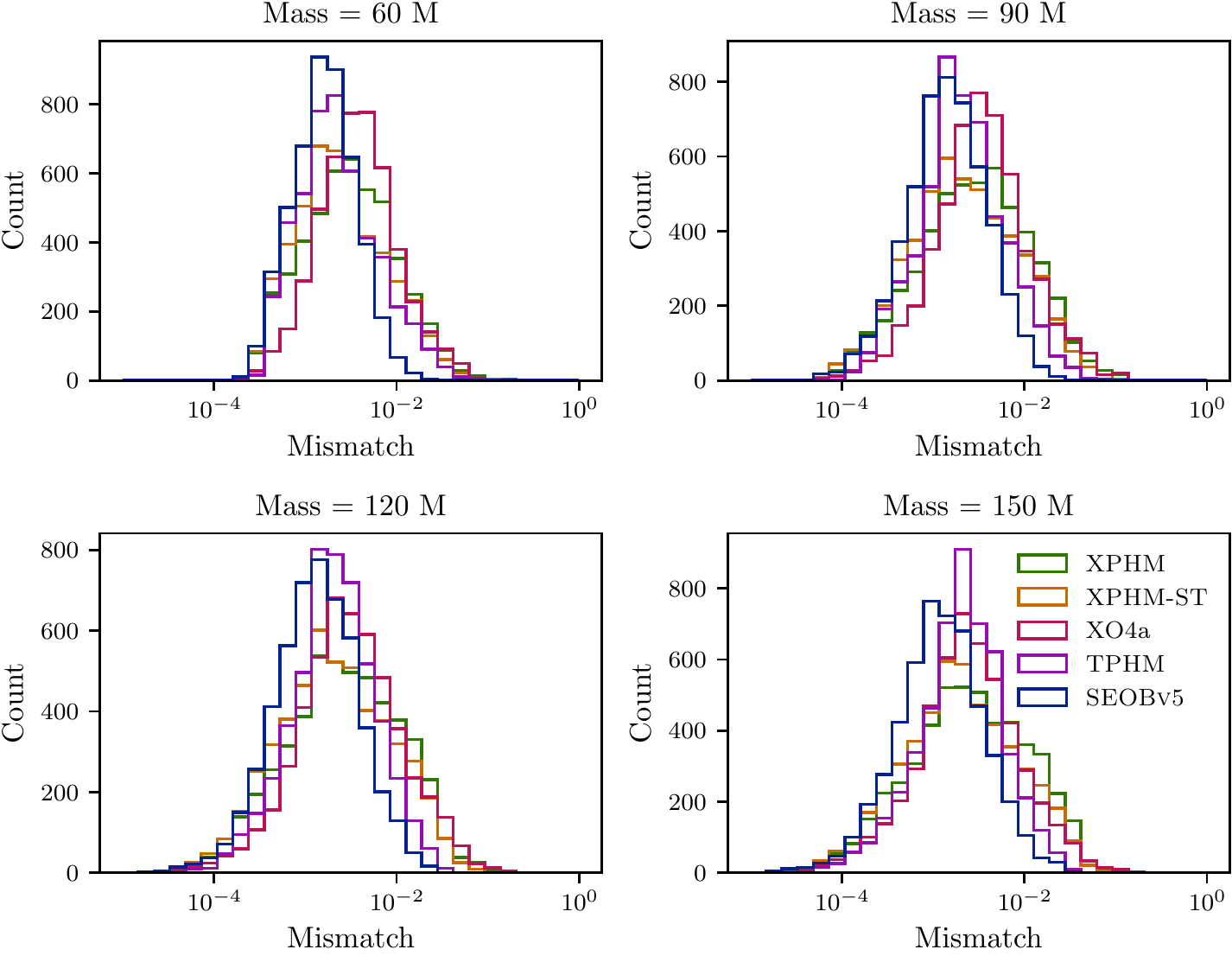}
   \caption{Mismatches between all multipoles up to $\ell=4$ in the $\mathbf{J}$ frame of  \nrsurSDQF{} and \modelname{} and \vFivePHM{}. Considered 1024 two-spin configurations at four different total masses $\{60,90,120,150\}M_\odot$ and five different inclinations $\{0,\pi/3,\pi/2,2\pi/3,\pi\}$ rad.}
   \label{fig: NRSur all 2spin}
\end{figure*}

To explore the $q\le4$ parameter space more thoroughly, we also consider the 1024 configurations generated using 
\nrsurSDQF{}. 
The results of this study is shown in Fig.~\ref{fig: NRSur all 2spin}, from which we can see that for the systems considered in this study, \vFivePHM{} performs best, followed by \tphm{}. The frequency-domain models \pxphm{}, with both the MSA and SpinTaylor angles, and \modelname{} have a tail to larger mismatches which degrade the overall performance of the model.
It is our understanding, informed by comparison with Fig.~\ref{fig: NRSur ell2 2spin} and of the middle and bottom panels in Fig~\ref{fig: BAM match}, that this arises from the non-trivial nature of the relative phasing of the higher multipoles in the frequency domain, which has not yet been completely understood.
Once this issue has been resolved, further tuning to the higher-order multipoles and extending the calibration to two-spin systems is expected to reproduce the improvement seen in the $(2,2)$-only and $\ell=2$ multipoles. (We note that the symmetric $\ell=2$ mismatches shown in \PaperOne illustrate
that tuning each ingredient in a \phenom{} model can lead in principle to competitive accuracy to \nrsur{} models, but with far fewer input waveforms.)
From Fig.~\ref{fig: NRSur all 2spin} we can also see that the broadening of the histograms toward lower mismatches with increased total mass seen for the $\ell=2$ results is not as strong when we consider systems which include higher order multipoles.

We also consider an additional 216 single-spin cases with the spin placed on the large black hole, thus mimicking the data set against which the model was calibrated. A comparison of the performance of the model against the single-spin and two-spin cases shows that the tuning to single-spin systems is not currently the dominant source of error in \modelname{}, given the comparable performance presented.

\begin{figure*}[htbp]
   \centering
   \includegraphics[width=0.98\textwidth]{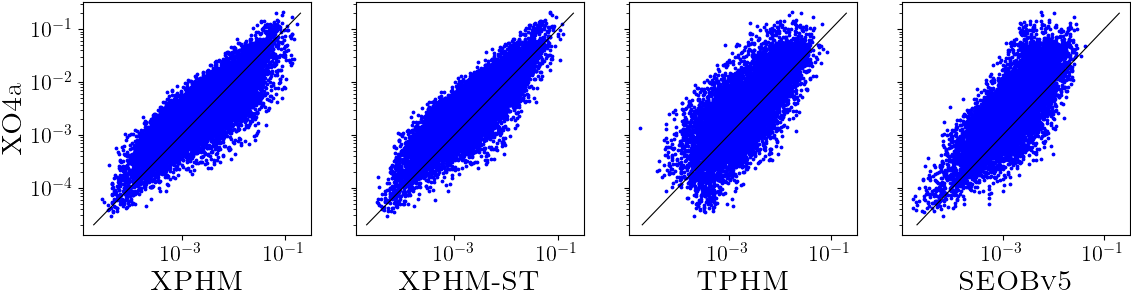}
   \caption{Mismatches between all multipoles up to $\ell=4$ in the $\mathbf{J}$ frame of  \textsc{NRSur7dq4} and selected models. Considered 1024 two-spin configurations at four different total masses $\{60,90,120,150\}M_\odot$ and five different inclinations $\{0,\pi/3,\pi/2,2\pi/3,\pi\}$ rad.}
   \label{fig: NRSur comparison}
\end{figure*}

An alternative comparison of the performance of \modelname{} against the other models considered in this study can be seen in Fig.~\ref{fig: NRSur 
comparison}. This shows that the time domain models \tphm{} and \vFivePHM{} generally perform better than the frequency domain models, which 
have a roughly comparable performance. However, the improvement does not exceed a factor of $\sim$2 at best. This is consistent with our 
expectation of the importance of correctly modelling relative phase offsets between multipoles, which is likely incorporated more naturally
in time-domain models, where the relative phases are inherited from the PN/EOB approximants.

We also considered the dependence of the mismatch on the inclination of the system.
The inclination here is measured with respect to the orbital plane of the binary at the reference frequency.
Since these are precessing systems, this will change throughout the evolution of the binary.
However, since these systems are uniformly distributed with $q\le4$ $|\chi_{1,2}|<0.8$ we do not 
expect the precession effects to be particularly strong for the majority of these systems --- for 
example, we do not expect to see transitional precession which would cause a strong change in the 
orientation of the binary during its evolution.

We observe that the spread of mismatch values is greatest for 
systems that are face-on and face-off at the reference frequency, i.e., both the best and 
the worst mismatches are seen here. This is true for all models.
It is for binaries which are originally edge-on where we see the greatest 
difference in model performance, with \vFivePHM{} showing a clear shift towards lower mismatch 
values compared to the other models, while for originally face-on/off systems the 
performance of all models is much more comparable for the bulk of systems, though 
the \textsc{Phenom} models have a noticeable tail to higher mismatches.

\begin{figure}[htbp]
   \centering
   \includegraphics[width=0.48\textwidth]{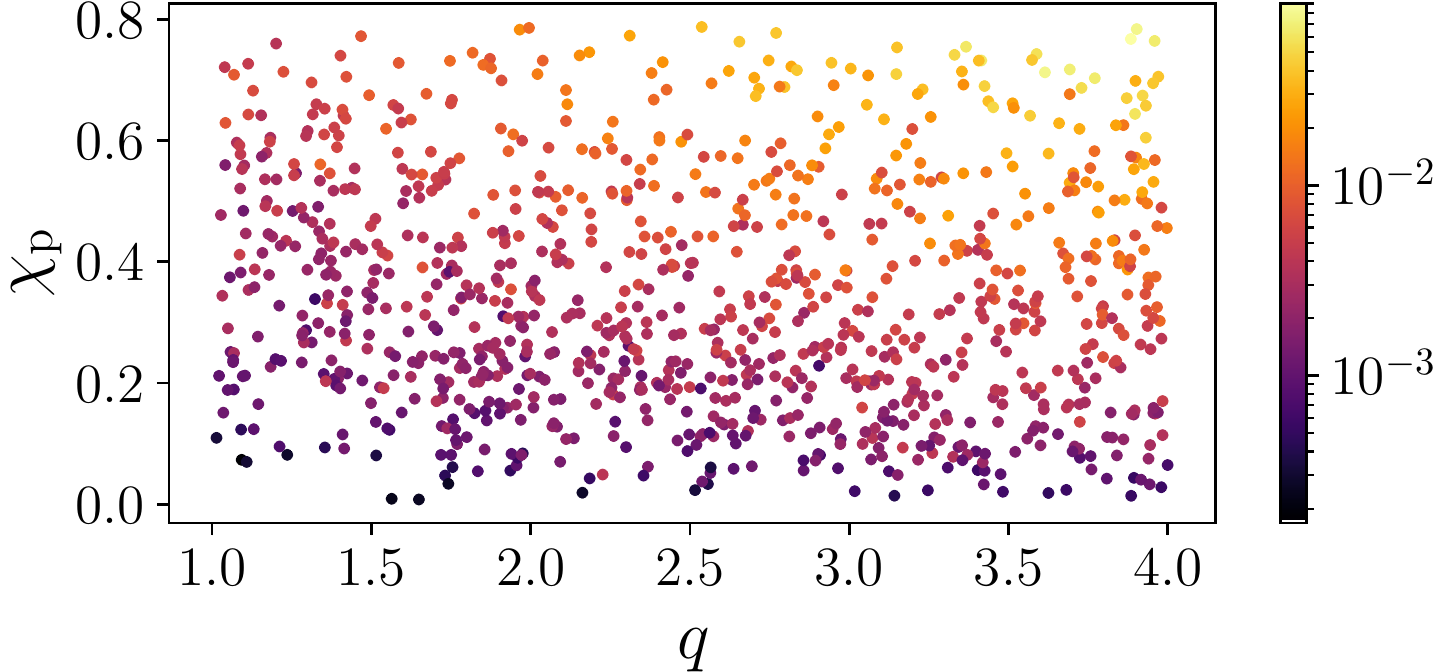}
   \caption{Dependence of the mismatches between all multipoles up to $\ell=4$ in the $\mathbf{J}$ frame of \nrsurSDQF{} and \modelname{} on $q$ and $\chi_\mathrm{p}$. Total mass $90M_\odot$, averaged over inclination.}
   \label{fig: mismatch q chip dependence}
\end{figure}

\begin{figure}[htbp]
   \centering
   \includegraphics[width=0.48\textwidth]{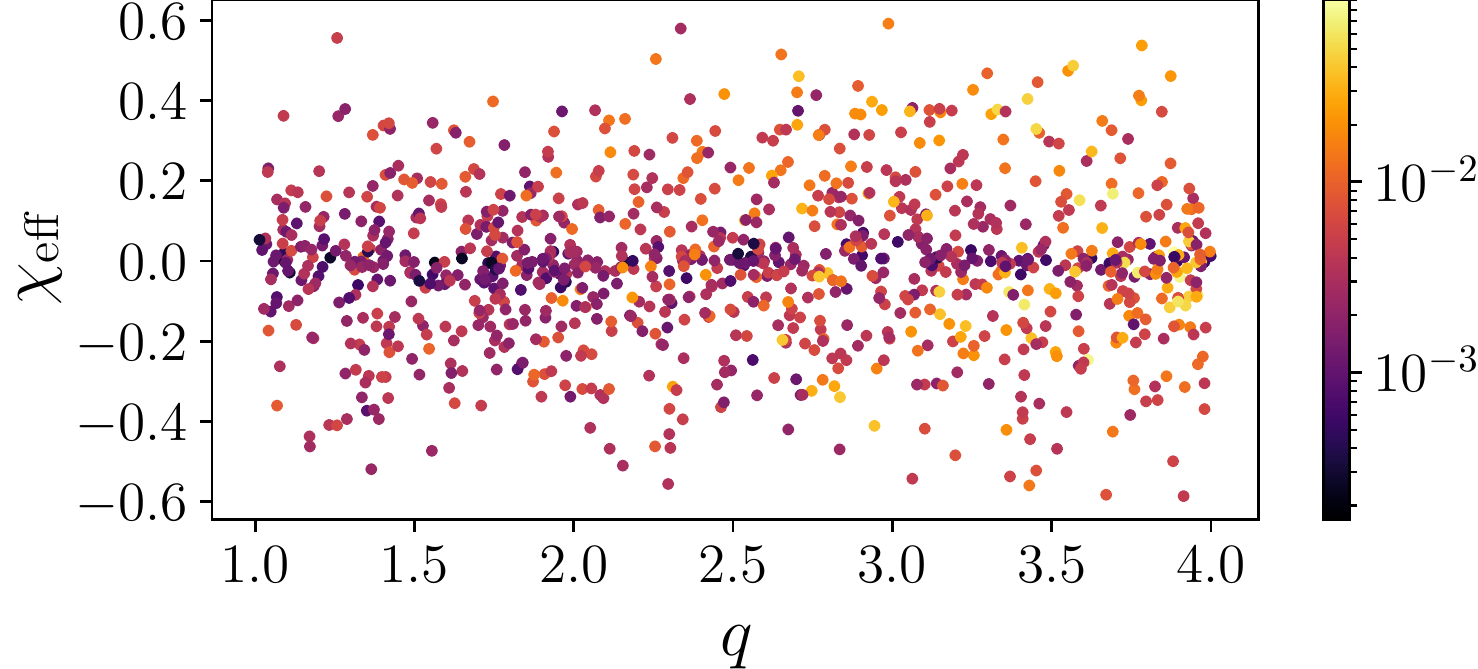}
   \caption{Dependence of the mismatches between all multipoles up to $\ell=4$ in the $\mathbf{J}$ frame of \nrsurSDQF{} and \modelname{} on $q$ and $\chi_\mathrm{eff}$. Total mass $90M_\odot$, averaged over inclination.}
   \label{fig: mismatch q chieff dependence}
\end{figure}

Finally, we consider how the performance of \modelname{} varies across the parameter space. The dependence on the mass ratio and dominant spin effects is demonstrated in Figs.~\ref{fig: mismatch q chip dependence} and~\ref{fig: mismatch q chieff dependence}. From this we can clearly see that \modelname{} performs best at low mass ratios and low in-plane spin values, as would be expected. The worst mismatches occur for systems with in-plane spin magnitudes above $\sim$0.6, with the very worst of these seen for systems with $q>3$.
The dependence on $\chi_\mathrm{eff}$ is much less strong, with poorer mismatches seen at all values of $\chi_\mathrm{eff}$. The best mismatches predominantly occur for systems with $|\chi_\mathrm{eff}|<0.1$.

\subsection{Parameter Estimation Results}
\label{sec:pe}
The mismatch study presented in Sec.~\ref{sec:full-matches} describes the accuracy of \modelname{} for single points in the parameter space. 
Although it was found that \modelname{} performs best at low mass ratios and low in-plane spin values, it only gives limited insight into how 
\modelname{} performs for realistic GW applications -- for example inferring the properties of the binary through Bayesian inference, 
e.g.~\cite{Veitch:2014wba}, and identifying GWs in noise, e.g.~\cite{Allen:2005fk,Babak:2012zx}. Previous attempts to correlate the mismatch 
with the model's performance for GW applications led to the indistinguishability criterion~\cite{Baird:2012cu}. Here, the mismatch was shown to 
relate to an \emph{indistiguishability SNR}, where biases in parameter estimates may be observed for GWs with SNRs larger than the
indistinguishability SNR. We therefore assess the accuracy of \modelname{} for realistic GW applications by performing Bayesian inference on 
both synthetic and real GW signals, with our choice of synthetic signal informed by identifying configurations where the SNR is above the 
indistinguishability SNR for all models. To quantify possible systematic biases, we compare the posterior distribution obtained with 
\modelname{} with a) the distributions obtained with \pxphm{}, \spintaylor{}, \tphm{} and \vFivePHM{}, 
and b) the true source parameters when they are known. We note that although a typical Bayesian analysis compares $\sim 10^{7}$ waveforms 
over a wide parameter space, we still only consider isolated GW signals. This means that we still only gain limited insight into the overall 
performance of \modelname{} over the full 15-dimensional parameter space.

The first two synthetic injections considered are generated with \nrsurSDQF{} as it most accurately resembles numerical relativity across its 
calibrated parameter space~\cite{Varma:2019csw}. We injected these signals into zero noise, which reflects the expected distribution when 
averaged over many different noise realizations. We use the expected detector sensitivities for the advanced LIGO and advanced Virgo GW detectors~\cite{LIGOScientific:2014pky,acernese2014advanced}. For all cases, we perform Bayesian inference with the {\textsc{dynesty}} 
nested sampler~\cite{Speagle:2019ivv}, employed via the {\textsc{Bilby}} parameter estimation 
software~\cite{Ashton:2018jfp,Romero-Shaw:2020owr}. All analyses used 1000 live points along with the bilby-implemented rwalk sampling algorithm with an average of 60 accepted steps per MCMC. We consistently used uninformative and wide priors, as defined in Appendix B.1 of Ref.~\cite{LIGOScientific:2018mvr}.

\begin{figure*}[htbp]
   \centering
   \includegraphics[width=0.49\textwidth]{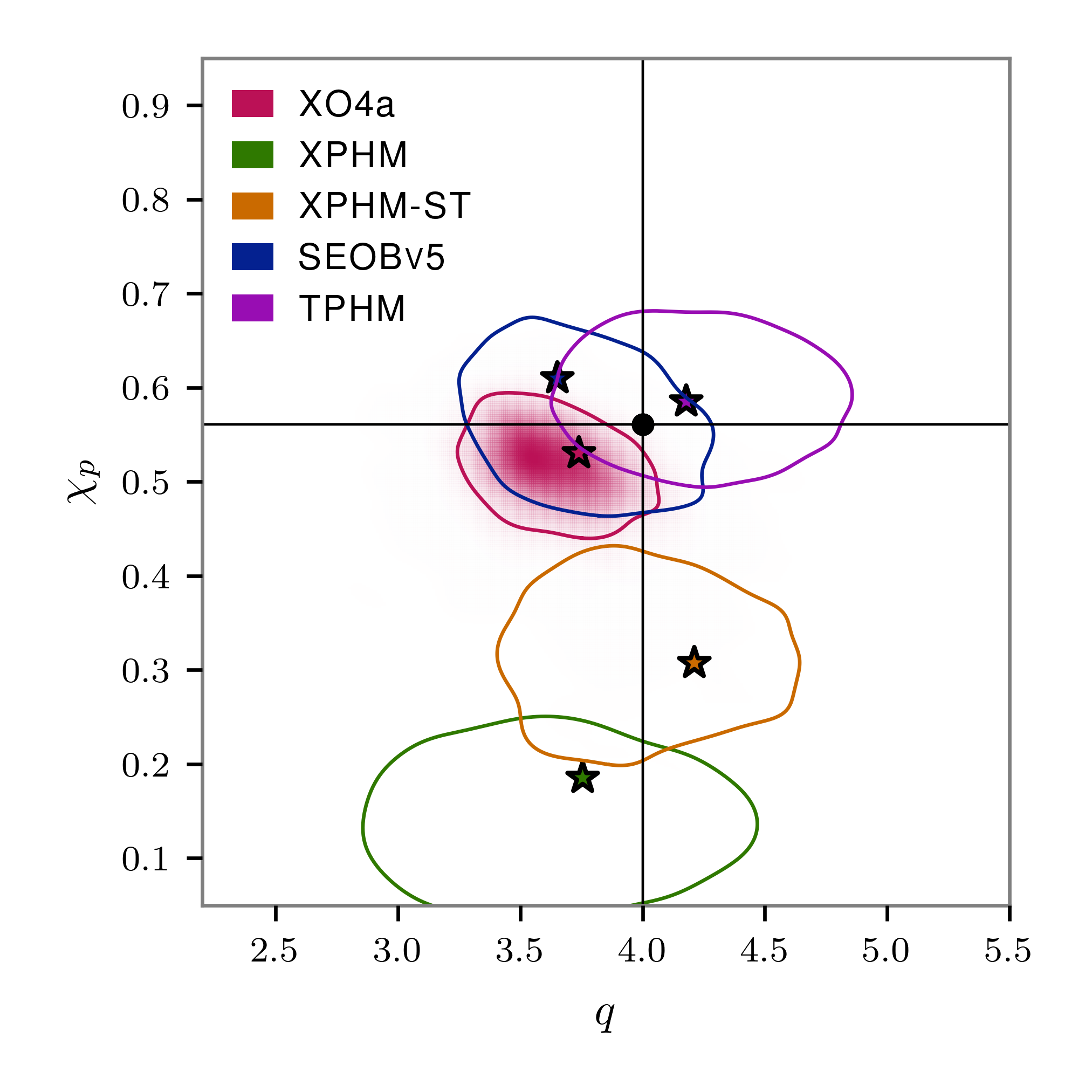}
   \includegraphics[width=0.49\textwidth]{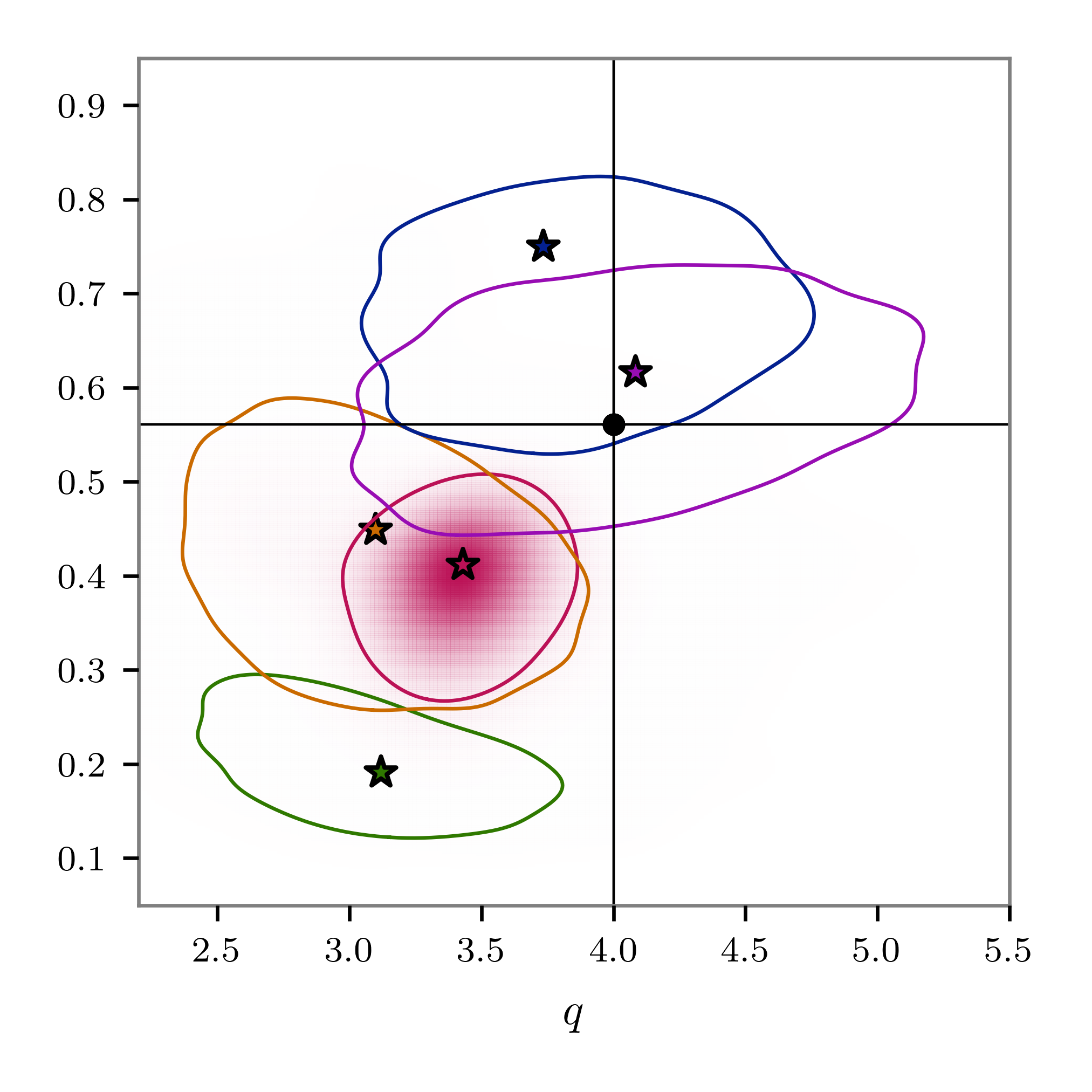}
   \caption{The two-dimensional marginalized posterior distributions for the mass ratio $q$ and effective precessing spin $\chi_{\mathrm{p}}$ when analyzing two gravitational wave signals simulated with \nrsurSDQF{}: \emph{Left}: low mass and \emph{Right}: high mass. The indicated two-dimensional area shows the inferred 90\% credible region, the horizontal and vertical black lines show the injected mass ratio and effective precessing spin, and the markers show the maximum likelihood position. The shaded region indicates the posterior probability density per pixel obtained with \modelname{}. The low and high mass simulated signals were produced for a binary black system with mass ratios $q=4$, effective spins $\chi_{\mathrm{eff}} = -0.45$, effective precessing spins $\chi_{\mathrm{p}} = 0.53$, and total masses $M=60\, M_{\odot}$ and $M=120\, M_{\odot}$ respectively. The distance to the source was chosen so that the simulated signals had signal-to-noise ratio $\mathrm{SNR}=20$.
   }
   \label{fig:pe_injection_1}
\end{figure*}

First, we analyze a synthetic GW signal for a fiducial binary black hole system with total mass $M=60\, M_{\odot}$, mass ratio $q=4$ and 
dimensionless spin vectors\footnote{The spin vectors are defined at $18.66\,\mathrm{Hz}$ with respect to the orbital angular momentum $\hat{\mathbf{L}}$} 
 $\chi_1 = [0.31, 0.47, -0.57]$ and $\chi_2 = [0.37, 0.71, 0.01]$. The effective
spin parameters are $\chi_{\rm eff} = -0.45$ and $\chi_{\rm p} = 0.53$. 
The inclination angle of the binary is set to $\iota=0.98\, \mathrm{rad}$, to emphasize the effect of higher-order multipoles and precession in the signal, 
and the luminosity distance is chosen to ensure a network SNR of $20$. All other extrinsic parameters are randomly chosen. We perform Bayesian 
inference on this specific binary configuration as we found a large variance in the match between \nrsurSDQF{} and each of the models 
used in this study at the true source parameters. This is therefore a suitable case to investigate how the matches presented in 
Sec~\ref{sec:full-matches} translate to performance with Bayesian inference. The mismatches at the true source parameters between \nrsurSDQF{} and 
\pxphm{}, \spintaylor{}, \modelname{}, \tphm{} and \vFivePHM{} are 0.040, 0.036, 0.032, 0.018, 0.009 respectively. 
Based on these mismatch results  we would expect only \vFivePHM{} to be indistinguishable from the injection with 90\% confidence; \vFivePHM{} is the 
only model with an indistinguishable SNR greater than the injected value\footnote{A quasi-circular binary black hole system has eight physical degrees 
of freedom: the two masses and 6 components of each spin vector. With eight degrees of freedom, the indistinguishable SNR is 
$\rho = \sqrt{6.68 / \mathfrak{M}}$ where $\mathfrak{M}$ is the mismatch.}. This means that we would expect \vFivePHM{} to recover the injected 
values most accurately, with possible biases in the other models.

In Fig~\ref{fig:pe_injection_1} we show the inferred two-dimensional marginalized posterior distribution for the mass ratio $q$ and effective precessing 
spin $\chi_{\mathrm{p}}$ with contours showing the inferred two-dimensional 90\% confidence interval (hereafter simply referred to as the 90\% 
confidence interval unless otherwise stated) and maximum likelihood samples. Although we see the general trend that a model with a larger match 
more accurately recovers the injected value, the trend is not trivial. For instance, although the mismatches for \modelname{} and both variants of 
\pxphm{} are comparable and much lower than \vFivePHM{}, the inferred 90\% confidence intervals are distinctly different. We see that the inferred 
distribution obtained with \modelname{} is comparable to \vFivePHM{} despite the mismatch being $\sim 3.5\times$ larger. Although all models 
prefer $q\sim 4$, only \modelname{}, \vFivePHM{} and \tphm{} prefer large in-plane spins; both variants of \pxphm{} show significant biases in the 
inferred $\chi_{\mathrm{p}}$ with the injected value lying significantly outside of the 90\% confidence interval. Although the injected value is outside 
the 90\% confidence region of \modelname{}, we see that the maximum likelihood samples for \modelname{}, \vFivePHM{} and \tphm{} are similarly 
spaced around the injected value.

Next, we investigate \modelname{}'s performance when the total mass of the fiducial binary black hole is increased from 
$M=60\, M_{\odot}$ to $M=120\, M_{\odot}$\footnote{The low (high) mass fiducial binary black hole with total mass 
$M=60\, M_{\odot}$ ($M=120\, M_{\odot}$) has individual component masses $48\, M_{\odot}$ and $12\, M_{\odot}$ ($96\, M_{\odot}$ and 
$24\, M_{\odot}$).}. In Fig~\ref{fig:pe_injection_1} we see that \tphm{} recovers the injected values most accurately, with the injected parameters 
lying within the $50\%$ confidence interval. Judging solely on the inferred 90\% confidence interval, we see that \vFivePHM{} is the next best 
performing model, with the injected values lying within the 90\% confidence interval. All other models show more significant biases, with the 
injected values lying outside of the inferred posterior distribution. When inspecting the maximum likelihood positions, we see that \tphm{} recovers 
a value close to the injected system, while \vFivePHM{} and \modelname{} are similarly distributed. We note that the posterior obtained with 
\modelname{} is significantly tighter than the other models, which is why it lies outside the 90\% confidence region. Assuming that the 90\% c
onfidence interval scales linearly with SNR~\cite{Cutler:1994ys,Poisson:1995ef} (this approximation is valid in the strong-signal limit), we would 
expect \vFivePHM{} to no longer recover the injected value within 90\% confidence for a signal with SNR $\sim 22$. When comparing the 
posterior distributions for the low and high mass cases, we see that for all models the inferred posterior distributions widen. This is expected as 
more massive binaries merge at lower frequencies, meaning that they have significantly fewer cycles within sensitive region of the detectors and 
consequently less information to break well known degeneracies. We also see that only \modelname{} prefers lower in-plane spins, with all other 
models preferring larger spins than their lower mass counterparts.

In general, we have shown that we cannot rely solely on the point-by-point match results to conclude whether a given model is accurate enough to recover the source properties with Bayesian inference. This is particularly highlighted by the fact that although \modelname{}'s mismatch for the low mass synthetic injection is $\sim 3.5\times$ larger than \vFivePHM{}, the inferred posterior distribution is comparable. We stress that the matches presented in Sec~\ref{sec:full-matches} are essentially draws from the likelihood surface, and therefore a single point estimate of the true values is not sufficient to understand the full likelihood, and consequently the model's performance across the full parameter space. 
This issue may be solved in the future by identifying a more informative mapping between the mismatch and potential biases in parameter estimates, or by sampling over the waveform model and using a parameter space dependent prior that describes the match manifold~\cite{Hoy:2022tst}. While waveform systematics will likely be an issue for loud GW signals with large mass ratios and in-plane spins, where all models report relatively large mismatches to NR, current estimates for the underlying distribution of black holes implies that this will not be an issue for the majority of observed signals~\cite{KAGRA:2021duu}.

Although we have only shown results for two synthetic injections, we analysed a further 100 injections with \modelname{}. We randomly drew binary parameters from a given prior distribution, while ensuring that the duration of each signal is less than 4 seconds. We injected each synthetic signal into Gaussian noise colored by the expected detector sensitivities for the advanced LIGO and advanced Virgo GW detectors, and then performed Bayesian inference with \modelname{}. When comparing the inferred posterior distributions against the injected values on a population level,  we obtain the expectations from a Gaussian likelihood.

Finally, we investigate \modelname{}'s performance when analyzing real gravitational wave candidates. We analyzed GW150914\_095045~\cite{LIGOScientific:2016aoc,LIGOScientific:2016vlm}, GW190412\_053044 (hereafter GW190412)~\cite{LIGOScientific:2020stg} and GW190814\_211039~\cite{LIGOScientific:2020zkf}. We focus our attention on GW190412 as the models used in the original LIGO-Virgo-KAGRA analysis~\cite{Khan:2019kot,Ossokine:2020kjp} gave conflicting parameter estimates; this is most strikingly shown in the inferred mass ratio and effective aligned spin posterior distributions, see Fig.~2 in~\cite{LIGOScientific:2020stg}. Although not discussed in detail, we note that the inferred posteriors for GW150914\_095045 and GW190814\_211039 were recovered as expected, and we refer the reader to Refs.~\cite{Colleoni:2020tgc,Islam:2020reh} for additional analyses of GW190412 which investigated potential model systematics. 

GW190412 was the first GW event observed by the LIGO-Virgo-KAGRA collaboration with confidently unequal component masses, evidence for the $(\ell, |m|) = (3, 3)$ multipole and marginal evidence for precession~\cite{LIGOScientific:2020stg,Hoy:2021dqg,Colleoni:2020tgc,Islam:2020reh}. Consequently, models with more accurate precession dynamics will give a more accurate reflection of the multipole structure of the observed GW and will more likely return unbiased parameter estimates.

\begin{figure}[htbp]
   \centering
   \includegraphics[width=0.49\textwidth]{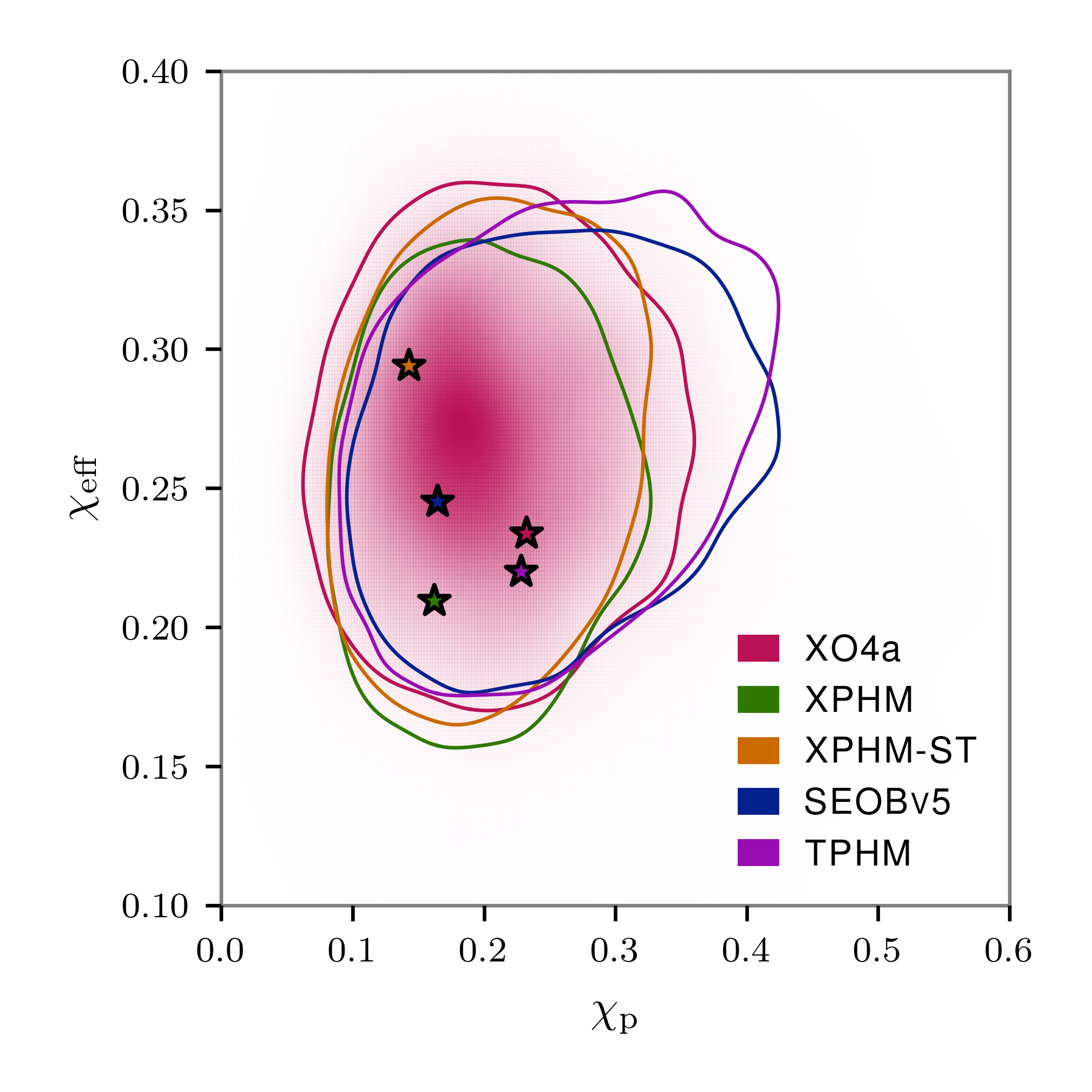}
   \caption{The two-dimensional marginalized posterior distributions for the effective precessing and aligned spin, $\chi_{\mathrm{p}}$ and $\chi_{\mathrm{eff}}$ respectively, when analyzing GW190412. The indicated two-dimensional area shows the inferred 90\% credible region and the markers show the maximum likelihood position. The shaded region indicates the posterior probability density per pixel obtained with \modelname{}.}
   \label{fig:pe_190412}
\end{figure}

To analyze GW190412, we use the strain data released by the Gravitational Wave Open Science Center (GWOSC)~\cite{LIGOScientific:2019lzm} and the publicly released power spectral densities and calibration envelopes included in the GWTC-2.1 data release~\cite{LIGOScientific:2021usb,ligo_scientific_collaboration_and_virgo_2022_6513631}. Although the original LIGO-Virgo-KAGRA analysis employed a highly parallelized version of \textsc{Bilby}~\cite{Smith:2019ucc}, we use the standard configuration, as described above. All settings were chosen to match the original LIGO-Virgo-KAGRA analysis.

In Fig~\ref{fig:pe_190412} we show the inferred two-dimensional marginalized posterior distribution for the effective precessing and aligned spin, $\chi_{\mathrm{p}}$ and $\chi_{\mathrm{eff}}$ respectively. We see that although \vFivePHM{} and \tphm{} recover a marginally tighter posterior for $\chi_{\mathrm{eff}}$ with slightly more support for larger $\chi_{\mathrm{p}}$, in general all models show excellent agreement for the bulk of the posterior. When comparing maximum likelihood samples, we see that all models approximately agree, but we note that \modelname{}  and \tphm{} prefer more precession, $\chi_{\mathrm{p}}\sim 0.23$ compared to other models $\chi_{\mathrm{p}}\sim 0.15$, and  \pxphm{} with the Spin Taylor precession angles prefers larger aligned spins. Importantly, the two-dimensional 90\% contours contain the maximum likelihood samples for all models.

In a typical GW Bayesian analysis, the likelihood is evaluated $\sim 10^{7}$ times. Having a waveform model that not only accurately describes the GW emitted from a binary merger, but also quick to evaluate, is therefore crucial for GW analyses. We find that the full \modelname{} waveform can be faster to evaluate close to equal mass, and takes longer to evaluate at highly asymmetric masses where the period of two-spin oscillations becomes more comparable to the orbital period. At worst the waveform generation is a factor of \(\sim \! 2\) slower than \pxphm{}. To check likelihood evaluation time we randomly drew $10^{6}$ points from a standard prior distribution and evaluated the likelihood for each point. We find that on average \modelname{} is $1.6\times$ and $1.2\times$ slower than \pxphm{} and \pxphm{} with the Spin Taylor precession angles respectively, $8.3\times$ faster than \vFivePHM{} and $3.4\times$ faster than \tphm{}. More information about waveform timing is available from the authors
upon request.

\section{Conclusions}
\label{sec:conclusions}

To date, analytic or semi-analytic binary-black-hole waveform models available for analysis of LVK observations have not included any
input from fully general-relativistic results (i.e., \nr{} simulations) for precession effects through the merger and ringdown, 
or the antisymmetric contribution to the signal multipoles. 
A model that included \nr{} tuning for the dominant symmetric $\ell=2$ contribution, \pnr, was presented in 
\PaperOne\!. For this model to be used in analysis of real GW signals, it would ideally 
be extended to include sub-dominant multipoles. In lieu of a full \nr{} tuning to the higher multipoles, the 
purpose of this work was to incorporate the precision dominant-multipole modeling of \pnr{} into a
state-of-the-art higher-multipole model infrastructure, \pxphm{}, and to apply an approximate
extension of the merger-ringdown precession dynamics to higher multipoles, and include a model for
the antisymmetric contribution~\cite{Ghosh:2023mhc}, to make a series of recent advances in precession
modeling~\cite{Hamilton:2021pkf,Hamilton:2023znn,Ghosh:2023mhc} available for GW signal analysis. 

In this first extension of the \pnr{} approach to higher multipoles, we have \emph{not} performed additional tuning 
to the higher multipoles, but have instead applied an approximate extension of the (2,2) results by (1) applying a model 
of the effective ringdown frequency to the coprecessing-frame higher multipoles, as described in Ref.~\cite{Hamilton:2023znn} 
and Sec.~\ref{sec:hm-coprec}, and (2) extending the frequency-domain precession angles for the higher
multpoles beyond the SPA prescription used in earlier models, to apply a frequency 
shift (rather than scaling) through the merger and ringdown, as described in Sec.~\ref{sec:hm-angles}.
In addition, we have incorporated a model of the antisymmetric contribution to the coprecessing-frame
(2,2) multipoles, as described in Ref.~\cite{Ghosh:2023mhc} and Sec.~\ref{sec:asymmetry}. We have also
made several improvements to the tuning of the precession angles, and, given the smooth behavior of
the \pnr{} model across the tuning parameter space, have retuned the model to use both the 40 
calibration waveforms used in the original \pnr{} model, plus an additional 40 waveforms that were 
previously used for validation; all \nr{} waveforms are publicly available and discussed in Ref.~\cite{Hamilton:2023qkv}.

As in the original \pnr{} model, we find a significant improvement in accuracy over all other models, if we
consider only the symmetric (2,2) contribution to the coprecessing frame. 
If we also consider the antisymmetric contribution in the coprecessing frame then, as we might expect,
 the accuracy of all symmetric models degrades. Our inclusion of a model for the antisymmetric contribution
 restores much of the accuracy. 
 
For the full precessing model, the accuracy is limited by the higher multipoles, and 
we find that the overall accuracy of all of the state-of-the-art models that we consider is comparable; in some
measures \modelname{} does slightly better than others (the \nr{} comparison in Fig.~\ref{fig: BAM match}, which extends up to $q=8$), 
while in other measures the \vFivePHM{} model is more accurate (e.g., mismatch 
histograms as shown in Figs.~\ref{fig: NRSur all 2spin}). We expect that the 
most significant error in the \modelname{} higher multipoles is in the relative time and phase offsets between
the different multipoles, which we have not explicitly checked or modeled. This may be a problem common to 
all frequency-domain models, since we find in parameter-estimation tests that the time-domain models 
\vFivePHM{} and \tphm{} show less parameter bias, despite including no precession tuning 
through merger and ringdown, and (in the case of \tphm{}) no clear advantage in 
mismatch comparisons. 

As noted in Sec.~\ref{sec:frame-choices}, some of the advantages of the ``twist-up'' procedure of modeling 
precessing-binary signals are lost when we move beyond the dominant multipole in the frequency domain. 
If we consider the frequency-domain multipoles in the QA frame, these will neither approximate those from a 
non-precesing binary in the inspiral, nor exhibit the simple structure of non-precessing-binary multipoles. 
Convsersely, if we wish to use non-precessing-binary multipoles in the coprecessing frame (as we do here)
then we do not have a well-defined prescription for the precession angles. In this work we have been guided
by the precession angles we calculate from considering only the $\ell=2$ multipoles, then the $\ell=3$ multipoles,
and so on. We do not have a procedure to determine the ``correct'' angles, against which we could check our
model. In future one must either find a way around these issues in order to produce accurate frequency-domain
models. 

Assuming that there is a way to move beyond these fundamental issues in frequency-domain modeling, 
\modelname{} needs to be improved in several ways: we need to model two-spin effects and the relative time and phase shifts 
between the higher multipoles. We also see that even the 
symmetric (2,2) coprecessing-frame modeling needs to be improved for configurations with high mass
ratios and high spins. As noted in \PaperOne\!, besides potential improvements in the 
ans\"atze to model the precession angles, one may also need to introduce an intermediate frequency 
region (and make use of longer \nr{} waveforms), and model the evolution of the direction of $\hat{\mathbf{J}}$. 

If we look more generally at the suite of state-of-the-art waveform models, we are less concerned with 
which is most accurate; the differences in accuracy vary significantly across the parameter space, and no
model is significantly and consistently more accurate than any other. Of far more interest is that \emph{none} 
are sufficiently accurate for upcoming observations. If we make the conservative accuracy requirement
that the mismatch must be lower than $\sim$$1/\rho^2$ for an SNR of $\rho$~\cite{Baird:2012cu}, then at 
$\rho=30$ we require a mismatch accuracy below $\sim$$10^{-3}$, and at $\rho=40$ we require mismatch 
uncertainties below $\sim$$10^{-4}$. This is a conservative estimate, but given that the typical mismatch
uncertainty is above $10^{-3}$ for \emph{all} precessing-binary models, a dramatic increase in accuracy is needed. 
The possible exceptions are
the NR surrogate models, but these are currently limited to high-mass binaries. As such it is possible that
systematic errors due to model innacuracy will be an issue for the loudest (and therefore most interesting)
signals in O4, and almost certainly in future observing runs, unless there is a significant (i.e., order of
magnitude) increase in overall accuracy of the state-of-the-art models. 

We also require a more complete understanding of the uncertainties in \nr{} waveforms, 
both as input to models as for validation, and a more robust methods to quantify the impact of model
uncertainties on parameter measurements. It is well known that the indistinguishability SNR is a 
conservative measure of waveform accuracy, and this is illustrated in Sec.~\ref{sec:pe}, where we
see that mismatch error of each model at the true parameters does not reflect the relative performance 
of each model in parameter recovery. Given that model accuracy is likely to be an important issue for
the most interesting observations, there is an urgent need for better methods to quantify the measurement
uncertainties that will result from modelling errors.

\section{Acknowledgements}

We the authors of this work would like to thank Angela Borschers, Rossella Gamba, 
Cecilio Garc\'ia-Quir\'os, Lucy Thomas, and Frank Ohme for their tireless efforts and assistance during the 
\modelnameLAL{} code review, and Tousif Islam for insighful comments about the manuscript. We also express thanks to Antoni Ramos-Buades , H\'ector Estell\'es, and Cecilio Garc\'ia-Quir\'os for their guidance in utilizing \texttt{pySEOBNR}.

JT, SG, PK, CH and MH were supported in part by Science and Technology Facilities Council (STFC) grant ST/V00154X/1 and 
European Research Council (ERC) Consolidator Grant 647839.
JT also acknowledges support from the NASA LISA Preparatory Science grant 20-LPS20-0005.
EH was supported in part by Swiss National Science Foundation (SNSF) grant IZCOZ0-189876 and by the UZH Postdoc Grant (Forschungskredit).
LL was supported at King's College London by Royal Society University Research Grant {URF{\textbackslash}R1{\textbackslash}211451}; and at the University of Amsterdam by the GRAPPA Prize.
SG was also supported from the Max Planck Society’s Independent Research Group program.
PK was also supported by the GW consolidated grant: STFC grant ST/V005677/1.
CH thanks the UKRI Future Leaders Fellowship for support through the grant MR/T01881X/1.
EH was also supported in part by the Universitat de les Illes Balears (UIB); the Spanish Agencia Estatal de Investigaci\'{o}n grants PID2022-138626NB-I00, PID2019-106416GB-I00, RED2022-134204-E, RED2022-134411-T, funded by MCIN/AEI/10.13039/501100011033; the MCIN with funding from the European Union NextGenerationEU/PRTR (PRTR-C17.I1); Comunitat Auton\`{o}ma de les Illes Balears through the Direcci\'{o} General de Recerca, Innovació I Transformaci\'{o} Digital with funds from the Tourist Stay Tax Law (PDR2020/11 - ITS2017-006), the Conselleria d’Economia, Hisenda i Innovaci\'{o} grant numbers SINCO2022/18146 and SINCO2022/6719, co-financed by the European Union and FEDER Operational Program 2021-2027 of the Balearic Islands; the ``ERDF A way of making Europe''.

The catalogue of numerical simulations against which this model was calibrated were performed on the DiRAC@Durham facility, managed by the Institute for Computational Cosmology on behalf of the STFC DiRAC HPC Facility (www.dirac.ac.uk). The equipment was funded by BEIS capital funding via STFC capital grants ST/P002293/1 and ST/R002371/1, Durham University and STFC operations grant ST/R000832/1. In addition, several of the simulations used in this work were performed as part of an allocation graciously provided by Oracle to explore the use of our code on the Oracle Cloud Infrastructure.

The authors are additionally grateful for computational resources provided by the LIGO laboratory and supported by National Science Foundation Grants PHY-0757058 and PHY-0823459, which were used to perform the match comparisons presented in this paper.

Most parameter estimation analyses were performed on the Sciama High Performance Compute (HPC) cluster, which is supported by the ICG, SEPNet and the University of Portsmouth. Additional parameter estimation and further analyses were performed on the supercomputing facilities at Cardiff University operated by Advanced Research Computing at Cardiff (ARCCA) on behalf of the Cardiff Supercomputing Facility and the HPC Wales and Supercomputing Wales (SCW) projects. We acknowledge the support of the latter, which is part-funded by the European Regional Development Fund (ERDF) via the Welsh Government. In part the computational resources at Cardiff University were also supported by STFC grant ST/I006285/1.

This research has made use of data or software obtained from the Gravitational Wave Open Science Center (gwosc.org), a service of the LIGO Scientific Collaboration, the Virgo Collaboration, and KAGRA. This material is based upon work supported by NSF's LIGO Laboratory which is a major facility fully funded by the National Science Foundation, as well as the Science and Technology Facilities Council (STFC) of the United Kingdom, the Max-Planck-Society (MPS), and the State of Niedersachsen/Germany for support of the construction of Advanced LIGO and construction and operation of the GEO600 detector. Additional support for Advanced LIGO was provided by the Australian Research Council. Virgo is funded, through the European Gravitational Observatory (EGO), by the French Centre National de Recherche Scientifique (CNRS), the Italian Istituto Nazionale di Fisica Nucleare (INFN) and the Dutch Nikhef, with contributions by institutions from Belgium, Germany, Greece, Hungary, Ireland, Japan, Monaco, Poland, Portugal, Spain. KAGRA is supported by Ministry of Education, Culture, Sports, Science and Technology (MEXT), Japan Society for the Promotion of Science (JSPS) in Japan; National Research Foundation (NRF) and Ministry of Science and ICT (MSIT) in Korea; Academia Sinica (AS) and National Science and Technology Council (NSTC) in Taiwan.

\appendix
\section{\texttt{LALSuite} Waveform Implementation}
\label{sec:waveform-flags}
In this section we provide a brief description of the waveform parameters available for using in \modelname{} and the locations of the specific fit coefficients mentioned throughout the text. 

In Table~\ref{tab:cp-pars} we present the tuning parameters for the coprecessing tuning 
outlined in Sec.~\ref{sec:cp-22} and listed in Eq.~\eqref{eq:deviation variables}. These variables
are presented alongside their names in the 
\texttt{LAL} code and the associated \pxas{} parameter they modify.
The final values of these fit coefficients are available within the file 
\texttt{LALSimIMRPhenomX\_PNR\_deviations.c} in \texttt{LALSuite}~\cite{lalsuite}.
\begin{table}
   \begin{tabular}{l*{3}{c}r}
       \hline
       $u_k$ \hspace{6pt} & Name in \texttt{LAL} code \hspace{6pt} & \pxas{} parameter ($\lambda_k$) \hspace{2pt}
       \\
       \hline
       $u_0$ & \texttt{MU1} & \texttt{pAmp->v1RD}
       \\
       $u_1$ & \texttt{MU2} & \texttt{pAmp->gamma3}
       \\
       $u_2$ & \texttt{MU3} & \texttt{V2}
       \\
       $u_5$ & \texttt{NU4} & \texttt{pPhase->cL}
       \\
       $u_6$ & \texttt{NU5} & \texttt{pWF->fRING}
       \\
       $u_7$ & \texttt{NU6} & \texttt{pWF->fDAMP}
       \\
       $u_8$ & \texttt{ZETA1} & \texttt{pPhase->b4}
       \\
       $u_9$ & \texttt{ZETA2} & \texttt{pPhase->b1}
       \vspace{2pt}
       \\
       \hline
   \end{tabular}

   \caption{Deviation variables for tuning of the $(\ell,|m|)=(2,2)$ multipole moment: Changes are made within \texttt{LALSimIMRPhenomX\_internals.c} and \texttt{LALSimIMRPhenomX\_precession.c} at or shortly after the definition of related structure parameters.  }
   \label{tab:cp-pars}
\end{table}

The fit coefficients for the tuned precession angles discussed in Sec.~\ref{sec:22angles-tuning}, 
along with the fit coefficients for the \(\beta_\text{f}\) fit detailed in Sec.~\ref{sec:final-beta-fit},
may be found in the file \texttt{LALSimIMRPhenomX\_PNR\_coefficients.c} in \texttt{LALSuite}. 
All results in this paper were produced with the coefficients found in the \texttt{git} hash: 08b90494b3a4a7dc966ca108e6c98eaa3f6e18a7.

In Table~\ref{tab:LAL-waveform-params} we show the available waveform parameter flags used to modify the 
behavior of \modelnameLAL{} present in \texttt{LALSimulation}, along with a description of the behavior 
that each parameter controls. One should note that, at present, enabling \texttt{PhenomXAntisymmetricWaveform} 
also requires enabling \texttt{PhenomXPNRUseTunedAngles} for the antisymmetric contributions to be present,
otherwise an error is raised upon waveform generation.

\begin{table*}
   \begin{tabular}{|l|l|c|c|}
      \hline
      Parameter name in \texttt{LAL} code & Description & Values & Default \\
      \hline
      \texttt{PhenomXPNRUseTunedAngles} & Enables tuned precession angles in \pxphm{} framework. & 0,1 & 0\\
      \texttt{PhenomXPNRUseTunedCoprec} & Enables tuned $(\ell,m)=(2,2)$ coprecessing model in \pxphm{} framework. & 0,1 & 0\\
      \texttt{PhenomXPNRInterpTolerance} & Sets the interpolation residual error threshold $R$ detailed in Eq.~\eqref{eq:delta-f-from-alpha}. & Float & 0.01\\
      \texttt{PhenomXAntisymmetricWaveform} & Enables the antisymmetric contribution to the $(\ell,m)=(2,\pm2)$ coprecessing multipole. & 0,1 & 0\\ \hline
   \end{tabular}

   \caption{The available waveform parameters added to \pxphm{} to activate the various 
   contributions that make up the \modelname{} model, coded in \texttt{LALSimulation} as \modelnameLAL. 
   Currently \texttt{PhenomXAntisymmetricWaveform} only activates when the flag is enabled and \texttt{PhenomXPNRUseTunedAngles} is also activated.}
   \label{tab:LAL-waveform-params}
\end{table*}

\bibliography{paper.bib}

\end{document}